%% file: changePartClassificationMining.tex
\PassOptionsToPackage{table}{xcolor}
\RequirePackage{fix-cm}
\documentclass[smallextended]{svjour3}       
\smartqed  
\usepackage{natbib}
\usepackage{graphicx}
\usepackage{listings}
\usepackage{xcolor}
\usepackage{colortbl}
\usepackage{tabularx}
\usepackage{multirow}
\usepackage{amsmath}
\usepackage{threeparttable}
\include{writing_support}

\definecolor{tableDark}{gray}{0.6}
\definecolor{tableMedium}{gray}{0.75}
\definecolor{tableLight}{gray}{0.85}

\newcommand{\rqAnswer}[2]{\begin{framed}\noindent\textbf{#1}: \textit{#2}\end{framed}}

\input{data}
\input{dataSzz}
\input{dataPerTicket}

\lstdefinelanguage{ruleset}{
    morestring=[b]',
    keywords = {and,or,skip,when,one,of,unless}
}
\lstdefinelanguage{pseudo}{
    keywords = {function,foreach,if,else,or,do,while,return},
    comment=[l]{//}
}

\journalname{Empirical Software Engineering}
\begin{document}

\title{An Industrial Case Study on Shrinking Code Review Changesets through Remark Prediction}

\titlerunning{Shrinking Code Review Changesets through Remark Prediction}

\author{Tobias~Baum \and
        Steffen~Herbold \and
        Kurt~Schneider
}

\institute{T. Baum, K. Schneider \at
              Fachgebiet Software Engineering \\
              Leibniz Universit\"at Hannover \\
              Hannover, Germany\\
              \email{\{firstname.lastname\}@inf.uni-hannover.de}           
           \and
           S. Herbold \at
              Universit\"at G\"ottingen \\
              G\"ottingen, Germany\\
              \email{herbold@cs.uni-goettingen.de}
}

\date{Received: date / Accepted: date}

\maketitle

\begin{abstract}
Change-based code review is used widely in industrial software development.
Thus, research on tools that help the reviewer to achieve better review performance can have a high impact.
We analyze one possibility to provide cognitive support for the reviewer: Determining the importance of change parts for review, specifically determining which parts of the code change can be left out from the review without harm.
To determine the importance of change parts, we extract data from software repositories and build prediction models for review remarks based on this data. The approach is discussed in detail. To gather the input data, we propose a novel algorithm to trace review remarks to their triggers. We apply our approach in a medium-sized software company.
In this company, we can avoid the review of 25\% of the change parts and of 23\% of the changed Java source code lines, while missing only about 1\% of the review remarks.
Still, we also observe severe limitations of the tried approach: Much of the savings are due to simple syntactic rules, noise in the data hampers the search for better prediction models, and some developers in the case company oppose the taken approach.
Besides the main results on the mining and prediction of triggers for review remarks, we contribute experiences with a novel, multi-objective and interactive rule mining approach. The anonymized dataset from the company is made available, as are the implementations for the devised algorithms.

\end{abstract}

\section{Introduction}
\label{sec:intro}

Code Review~\citep{fagan1976design}, especially in the form of ``change-based code review''~\citep{Rigby2013,baum2016qrs}, is a technique that is widely used in contemporary software development~\citep{baum2017profes}. Therefore, finding ways to improve the efficiency and effectiveness of reviewers is of high practical relevance. One of the main problems in change-based code reviews is dealing with large changesets~\citep{bacchelli2013expectations}. Code review tools that provide \emph{cognitive support} have been proposed as a means to reduce this problem and improve reviewer performance~\citep{baum2016profes}. One starting point for cognitive support is the observation that the parts of the changeset under review differ in importance. For example, consider a change that corrected the indentation of code in comparison to adding an algorithm. Intuitively, it is much more important to review the latter. Within this article, we build an empirical foundation for this intuition. We use repository mining to infer the importance of changes for reviews: Modern development processes usually generate vast amounts of review-related data from source code management systems (SCMs), ticket systems, and review tools \citep{Hamasaki2013,Bird,Thongtanunam2015}.

Thus, \emph{the purpose of this study is to analyze ways to use repository mining to identify the importance of change parts for code review and to improve code review efficiency based on this information.} We include a discussion on the foundations of our approach, rooted in research on code reviews and (to some degree) cognitive psychology. Furthermore, we discuss the technical issues of extracting the relevant data from repositories and mining knowledge from this data. We applied the resulting concepts in a case study at a medium-sized software company and report results from this case study. Our ambition is to go the whole way from the initial discussion to the use in practice. By using an interactive and multi-objective approach for rule mining, we left the beaten path followed by most of the current defect prediction research to explore an alternative we regard as promising.

We focus on support for the reviewer. With our first research question, we derive possibilities to help the reviewer, assuming that we know the importance of the parts of the code change under review:
\begin{examplebox}
\textbf{RQ$_{1}$.} How can information on change part importance help to reach code review goals more efficiently? What is ``importance'' in this regard?
\end{examplebox}
A model to determine change part importance can be built based on empirical data that shows which change parts act as triggers for review remarks. We motivate and propose an algorithm to mine this data from source code repositories. By analyzing it in contrast to the SZZ approach~\citep{Sliwerski2005} that is usually used in defect prediction studies, we show that they lead to different results:
\begin{examplebox}
\textbf{RQ$_{2.1}$.} How can potential triggers for review remarks be extracted from software development repositories?\\
\textbf{RQ$_{2.2}$.} How do the results of our extraction algorithm differ from those of the SZZ algorithm?
\end{examplebox}
Based on the gathered data, we build a prediction model. We choose a mining approach based on preferences articulated by the developers at our industrial partner. We evaluate the results of our approach and a standard rule mining approach, both subjectively with the developers and objectively based on performance metrics:
\begin{examplebox}
\textbf{RQ$_{3.1}$.} Which requirements for the classification model are considered most important by the developers in the case study company?\\
\textbf{RQ$_{3.2}$.} What are the characteristics of good rulesets found in the data?\\
\textbf{RQ$_{3.3}$.} How good are the found rulesets in the developers' opinions?\\
\textbf{RQ$_{3.4}$.} How well do the found rulesets perform on unseen data?
\end{examplebox}

Next, we outline the methodology and describe the case study context, before delving into the details of the study.

\section{Overall Methodology}
\label{sec:methodology}

In the following, we describe our overall methodology. Our work is a case study, as we look at review remark prediction in a single real-life context. The final goal is to make reviews in the company more efficient. Our research design is flexible, and we pragmatically mix methods: Parts of the study are deductive (\eg determining the goal of the mining in Section~\ref{sec:classificationImproveReview}, and parts of the feature selection in Section~\ref{sec:features}). Other parts, especially the data mining and the evaluation of the found rules, are empirical. We combine qualitative as well as quantitative data, \eg when triangulating the quality of the found rules from the results on the extracted data (Sections \ref{sec:ruleMiningResults}~and~\ref{sec:performanceUnseenData}) and opinions from the team (Section~\ref{sec:developersOpinion}). The specific data sources will be described in the sections where they are used.

Both case study research, as well as data mining, are highly iterative endeavors~\citep{runeson2012caseStudy,mariscal2010survey}. We also used an iterative approach, starting with the initial stages of the study in fall 2017, until the discussion of the study's results with the development team in fall 2018. The current report linearizes the results. It starts by describing the case study setting in detail (Section~\ref{sec:caseStudySetting}). Then, our detailed data mining choices will be discussed: The goal of the mining process (Section~\ref{sec:classificationImproveReview}), extraction of potential triggers for review remarks from repository data (Section~\ref{sec:howToExtract}), requirements and our approach for building a prediction model from the mined data (Section~\ref{sec:howToBuildModel}) and feature selection (Section~\ref{sec:features}). Finally, the results of applying our approach to the company data are shown in Section~\ref{sec:results}. We conclude by discussing problems, implications and future work (Section~\ref{sec:futureWork}), threats to validity (Section~\ref{sec:threats}) and related work (Section~\ref{sec:relatedWork}).

\section{Case Study Setting}
\label{sec:caseStudySetting}

In the following we characterize the case study setting in detail.
The company we collaborated with in this study is a medium-sized software company from Germany. It develops and sells a software product suite, as well as related consulting and services. Many of its customers are from the financial or public sector. The most severe consequence of defects can be financial losses for the customers. The company employs around 70 people, of which about 18 are full-time developers (14 in the year 2013, where our earliest empirical data is from). The study's first author also works part-time for the company. 

In the study, we analyze data from 2013 to 2018. 
At the start of this period, the SCM repository contained about 28,000 files and directories. Of these, about 7400 are Java files with a total of about 950,000 lines of Java code\footnote{simply counting lines, not distinguishing between empty lines, comments, etc}.
At the end of the timeframe, the repository contained about 86,000 files and directories, of which about 20,000 are Java files with about 2.3 million lines of Java code.
Java is the primary implementation language, only recently TypeScript was introduced as a second language for UI code.

The company's development team has been using an agile Scrum/Scrum-ban process since 2008. Currently, sprints last three weeks, and there is a public release of the product at the end of each sprint. There is a culture of collective code ownership, and the developers shall be able to work as generalists on many parts of the codebase. The team works mostly co-located. Code quality is important, and code style, unit tests, and many other quality checks are performed on a CI server. The development is done trunk-based in a monolithic Subversion repository. For this study, a git copy of the repository was used for ease of access and performance.

\begin{table}
\caption{The Company's Code Review Process (described along the relevant classification facets from \citet{baum2016qrs})}
\label{tab:codeReviewFacets}
\begin{tabularx}{\textwidth}{X}
\hline
\rowcolor{tableMedium}Unit of work = Task:\\
The team divides user stories into separate implementation tasks. A review is performed for each such task. There are separate ``bug'' and ``impediment'' (\ie internal improvement) change tasks that are not split but reviewed as a whole. All change tasks are hosted as tickets in Atlassian Jira.\vspace{1mm}\\ \hline
\rowcolor{tableMedium}Enforcement for triggering = Tool:\\ 
Separate states in the Jira ticket workflow ensure that a review candidate is created for each task. The relevant states are ``ready for review'', ``in review'', ``review rejected'' and ``done''.\vspace{1mm}\\ \hline
\rowcolor{tableMedium}Publicness of the reviewed code = Post commit:\\ 
A review is performed after the changes are visible for other developers, \ie after committing to the Subversion repository.\vspace{1mm}\\ \hline
\rowcolor{tableMedium}Number of reviewers:\\ 
There is usually one reviewer per ticket, but for specific modules two reviewers are mandatory. Pair programming reduces the number of needed reviewers by one.\vspace{1mm}\\ \hline
\rowcolor{tableMedium}Reviewer population = everybody:\\ 
Every team member shall be available as a reviewer for every change, albeit very inexperienced team members usually don't review alone.\vspace{1mm}\\ \hline
\rowcolor{tableMedium}Interaction while checking = on-demand:\\ 
The review participants only interact on-demand,\eg when there are questions regarding the code. Very rarely, reviews are performed in an in-person meeting.\vspace{1mm}\\ \hline
\rowcolor{tableMedium}Temporal arrangement of reviewers = sequential:\\ 
Only one reviewer reviews a ticket at a time and rework is done after each reviewer.\vspace{1mm}\\ \hline
\rowcolor{tableMedium}Detection aids = sometimes testing:\\ 
Sometimes, reviewers perform a limited amount of manual exploratory testing during reviews.\vspace{1mm}\\ \hline
\rowcolor{tableMedium}Reviewer changes code = sometimes:\\ 
The reviewers may change code during checking and commonly do so for minor changes.\vspace{1mm}\\ \hline
\rowcolor{tableMedium}Communication of issues = written:\\ 
The found issues are mainly communicated in writing and stored in Jira.\vspace{1mm}\\ \hline
\rowcolor{tableMedium}Options to handle issues = resolve, reject:\\ 
The author usually fixes observed issues right away or decides together with the reviewer that the remark will not be fixed. Consequently, review changes are almost always done in the same ticket as the original implementation.\vspace{1mm}\\ \hline
\rowcolor{tableMedium}Tool specialization = general-purpose, later specialized:\\ 
The traditional way of performing reviews in the company is by looking at the source code changes in TortoiseSVN, a graphical client for Subversion. In 2016 the company introduced a specialized, IDE-integrated code review tool (CoRT, http://github.com/tobiasbaum/reviewtool) that is now used by most of the developers.\\
\end{tabularx}
\end{table}

The company introduced regular code reviews in 2010, and the general process stayed stable over the last years.
For almost every development task (ticket), a review is performed for the corresponding code changes. When there are review remarks, these are usually fixed in further commits for the same ticket. Table~\ref{tab:codeReviewFacets} contains a detailed description of the code review process.

\section{How Can Change Part Classification Help to Reach Code Review Goals more Efficiently?}
\label{sec:classificationImproveReview}

\subsection{Possibilities for Using Change Part Importance to Improve Review}
\label{sec:possibilitiesChangePartImportance}

For many software developers, it is obvious that not all parts of an artifact under review are equally important for the review. We could observe corresponding statements in one of our recent studies~\citep{baum2017icsme}, but it has also been mentioned in much older works, \eg by \citet{Gilb1993}.
But what is ``importance''? Stated fully, it means ``importance for one of the review goals''. Code reviews are used for a variety of reasons~\citep{bacchelli2013expectations,baum2016fse}, of which finding defects and quality problems and spreading knowledge are usually among the most important ones. We will elaborate on the various review goals in Section~\ref{sec:targetMetric}.

Assuming that one knows the importance of a change part for review, a review tool can use this information in several ways\footnote{We explicitly focus on support for the reviewer during reviews. There are further ways to use repository mining to improve review outcomes, for example with automatic review agents (see Section~\ref{sec:relatedWork}).}:
\begin{enumerate}
\item Leaving out unimportant change parts
\item Highlighting very important change parts
\item Ordering the change parts from most important to least important
\end{enumerate}

Possibility~1 provides support in two ways: It saves time and effort directly because there is less to review. But we also expect an indirect effect: Less code to review means less cognitive load, and that in turn increases the chance of finding issues in the remaining change parts~\citep{baum2018orderingExperiment}.

The theoretical foundation for the positive effect of possibility~2 lies in the ``SRK taxonomy''\citep{rasmussen1983skills}, which states that human cognitive processing can happen in one of several modes. The skill- and rule-based modes depend to a large degree on automatic processing/pattern matching, whereas knowledge-based processing requires much more laborious conscious mental effort. By highlighting very important change parts, the tool can help the reviewer to choose the most efficient behavior from the SRK taxonomy. Parts where knowledge-based processing (i.e., thinking hard) is most likely to be beneficial can be highlighted, the rest of the change parts can be processed rule-based or skill-based and therefore more efficiently.

At first glance, it might seem that the classifiers underlying possibilities 1~and~2 differ just in the importance threshold used. But the difference lies deeper: The distinction in possibility~1 is between ``no processing'' and ``processing'', whereas the distinction in possibility~2 is between ``knowledge-based processing'' and ``rule-based processing''.

Possibility~3 (ordering by importance) provides cognitive support by making the reviewer aware of which parts are important and which are not. Thus, it combines aspects of possibilities 1 and 2. Furthermore, it ensures that the most important parts have been read when the review is ended prematurely. But it has weaknesses: It does not respect the logical structure of the change, so that the resulting order can be confusing to the reviewer~\citep{baum2017icsme}. Moreover, it does not provide a clue for the reviewer whether and when the review should be ended early.

All three possibilities can probably be put to good use for review support, but we will focus on possibility~1. It is in line with our backing theory of cognitive load, and we expect the highest potential for efficiency gains here. Also, it does not require modeling the difference between rule-based and knowledge-based processing and is, therefore, easier to operationalize than possibility~2.

\vspace{1cm}

\subsection{Influence of Leaving out Change Parts on Possible Review Goals and Derivation of Target Metrics}
\label{sec:targetMetric}

\begin{table}[b]
	\centering
	\caption{Assessment of the Effect of Leaving out Change Parts from Reviews on the Attainment of Review Goals and Avoidance of Unintended Review Side-Effects.}
	\label{tab:reviewGoalAssessment}
	\rowcolors{2}{tableLight}{white}
		\begin{tabular}{p{11cm}}\hline
\rowcolor{tableDark}\emph{Review goals:}\\
\textbf{Better code quality.} The direct positive effect of code reviews on code quality is based on the correction of the code according to the review remarks. Reaching this goal should be unaffected as long as the review of the reduced set of change parts still leads to the same fixed remarks.\\
\textbf{Finding defects.} Same as above, reaching this goal depends on the fixed remarks.\\
\textbf{Finding better solutions.} When the review discussion leads to better solutions, this results in changes to the codebase (\ie fixed review remarks), so an indicator that this goal is still reached is again when the fixed remarks stay the same.\\
\textbf{Learning (author).} The code's author mainly learns from the review remarks communicated to him/her, so that this goal will be unaffected when the communicated remarks stay the same.\\
\textbf{Learning (reviewer).} This goal is unaffected when the knowledge gained from the shrunk review set is comparable to the original review. A necessary (but not sufficient) precondition for that is that the shrunk change under review is still understandable.\\
\textbf{Complying to QA guidelines.} When the quality assurance guidelines demand the review of certain parts of the code, these may not be left out from the review scope.\\
\textbf{Improved sense of mutual responsibility.} The attainment of this goal mainly depends on doing reviews at all, not on the exact review scope.\\
\textbf{Team awareness.} The attainment of this goal mainly depends on how the reviews are communicated, not on the exact review scope.\\
\textbf{Track rationale.} Review remarks can point to portions of the code with an insufficient description of the rationale, so the attainment of this goal also depends on the set of communicated review remarks.\\
\textbf{Avoid build breaks.} Usually, build breaks can be avoided more efficiently by automatic checks than by manual reviews. In case reviews are still used for this goal, its attainment also depends on the set of fixed remarks.\\\hline

\rowcolor{tableDark}\emph{Review side-effects:}\\
\textbf{Avoid higher staff effort} Leaving out parts to review should usually lead to lower review effort and therefore help to attain this goal. A potentially adverse effect could occur when the understandability of the change is hampered by leaving out change parts needed for understanding.\\
\textbf{Avoid increased cycle time} Same as above, the review duration could only be negatively affected when the understandability of the change is hampered.\\
\textbf{Avoid offending/social problems} We see no potential that leaving out change parts based on objective rules could influence the risk of offending people or provoking other social problems through reviews. \\\hline
		\end{tabular}
\end{table}

In Table~\ref{tab:reviewGoalAssessment} we discuss for each of the common review goals how leaving out change parts can make a difference in achieving the goal. The goals have been consolidated from \citet{baum2016fse} and \citet{bacchelli2013expectations}.
The essence of the analysis is that most of the goals depend either on the communicated or the fixed review remarks. The decision whether to communicate a remark or to fix it on the fly is unlikely to be affected by shrinking the review scope so that we will treat these the same in the following.

We hypothesize that there are two main preconditions for a reviewer to find a remark: The reviewer needs to understand the relevant portions of the code and needs to observe a \emph{trigger} for the remarks. Consequently, the set of review remarks stays the same when the relevant code portions are still understandable and when there is still at least one trigger for each review remark in the shrunk change. The notion of triggers for review remarks will be discussed further in Section~\ref{sec:howToExtract}.

Two goals do not directly relate to review remarks: ``complying to QA guidelines'' and ``learning of the reviewer''. For many teams, these are the least important goals\footnote{In the survey by~\citet{baum2017profes}, the mode for ``complying to QA guidelines'' is ``last rank'' with 45\%, the mode for ``learning of the reviewer'' is ``rank before last'' with 28\%},
whereas ``better code quality'' and ``findings defects'' are very often among the top reasons for doing reviews~\citep{baum2017profes}. Therefore we will focus on shrinking the review scope while leaving the set of triggered remarks and the understandability of the change intact.

In commercial software development, code reviews are usually not an end in itself but are a means among several to obtain a high profit or return on investment. Therefore, it is often acceptable to find a few review remarks less when this is offset by higher savings in the review effort.

\begin{equation}
\label{eqn:profit}
\begin{split}
\mbox{profit}_{shrinking} &= \mbox{savings} - \mbox{effort}_{missedRemarks}\\
&= \sum_{l \in \textnormal{leftOut}} \mbox{reviewEffort}(l) - \sum_{m \in \textnormal{missedRemarks}} \mbox{cost}(m)
\end{split}
\end{equation}

As a simplification, we assume a constant cost factor $c$ that relates the review effort for change parts to the effort caused by missed remarks, \ie $\mathit{profit} \propto \left|\mathit{leftOut}\right| - c * \left|\mathit{missedRemarks}\right|$.
The time needed to review a change part can be estimated based on the empirical data from code review logs, but the effort caused by a missed review remark is much harder to quantify. We use two alternatives: (1)~The cost factor can be treated as unknown and the break-even point for a positive profit calculated, \ie the maximum cost per missed remark up to which the profit will still be positive for a given rule. A rule with a higher break-even point is more conservative and therefore usually better. (2)~We provide the user of the mining UI (see Section~\ref{sec:howToBuildModel}) with the results for several cost factors so that he or she can easily compare the possibilities.

\rqAnswer{RQ$_{1}$}{Information on change part importance can help the reviewer to reach code review goals more efficiently by: (1)~Leaving out change parts that neither trigger review remarks nor are needed for understanding, (2)~highlighting change parts that the reviewer should review in a higher cognitive mode than he or she otherwise would, or (3)~order change parts from the most likely to the least likely remark trigger, as long as this order is still easily understandable. We focus on (1) in the rest of the article and take the point of view that a slight decrease in found remarks is acceptable when offset by a significant saving in effort, \ie when the overall profit is positive.}

\section{Extracting Potential Triggers for Review Remarks from Repositories}
\label{sec:howToExtract}

\subsection{Remarks, Triggers, and Change Parts}
\label{sec:remarksAndTriggers}
\begin{figure}[b]
\centering
  \includegraphics[width=0.8\textwidth]{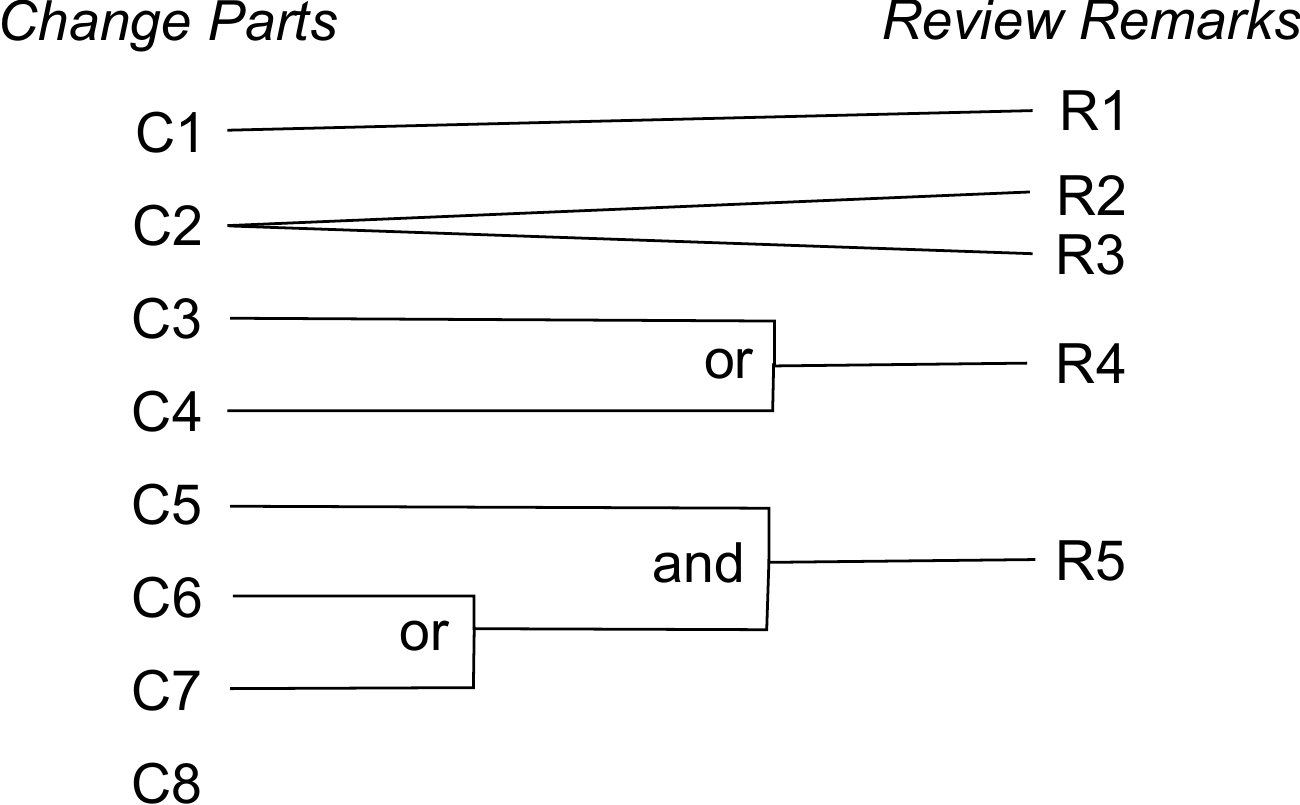}
\caption{Several possibilities how change parts can act as triggers for review remarks}
\label{fig:triggers}
\end{figure}

To judge the importance of a change part for review, we need to tell whether it contains triggers for review remarks. A trigger is the portion of code whose review leads to the creation of the review remark. In a simple case, the trigger is a defect in a specific line of the code. Another example could be a misspelled method name: Every occurrence of the method name in the code can lead to the observation of the problem and creation of the remark.
As this example shows, the relation between remarks, triggers, and change parts is not a simple one-to-one relation: There can be multiple types of composite conditions. Several of the possibilities are depicted in Figure~\ref{fig:triggers}.

The simplest possibility is when a change part acts as a trigger for exactly one review remark, as for C1 and R1. A single change part can also lead to several remarks (C2, R2, and R3). In some cases, there can be more than one trigger for a remark. One example is the typo in a method name mentioned above. This situation is similar to a logical ``or'': R4 is triggered when C3 or C4 is reviewed. In other cases, the knowledge of several parts of the code is needed to spot a problem and trigger a remark, \ie a logical ``and''. The figure shows a combined case, where reviewing C5 and at least one of C6 or C7 triggers the remark R5. Naturally, there are also change parts that do not act as a trigger at all (C8), which is the primary motivation for our current study.

The concept of ``and'' relations in triggers is related to the issue of understandability for the reviewer mentioned in the previous section. It is out of the scope of the current article.

\subsection{Selecting a Data Source}
\label{sec:selectingDataSource}

We will now discuss how the information on potential review triggers can be extracted from software repositories. Our technique is based on the assumption that one of the triggers for a review remark is close to the position of the remark.

\begin{table}[b]
\caption{Comparison of the Benefits of the two Options to Extract Review Remarks from Software Repositories}
\label{tab:discussionDataSource}
	\rowcolors{2}{tableLight}{white}
\begin{tabularx}{\textwidth}{X}
\hline
\rowcolor{tableDark}\emph{Benefits of extracting the fixed review remarks from the SCM:}\\
The remarks that led to changes in the codebase are arguably the most practically relevant.\\
They can be extracted quite easily and with high accuracy from the SCM, only ticket IDs and time intervals for the reviews are needed.\\
Gathering complete data is easy.\\
Remarks that have been communicated only orally will not be missed.\\
The technique is also usable in settings where review remarks are not stored in a structured form. \vspace{1mm}\\ \hline

\rowcolor{tableDark}\emph{Benefits of extracting the review remarks from the review tool database:}\\
The remarks that the reviewer took the time to write down textually are arguably the most practically relevant. (This definition of ``practical relevance'' would contradict the one given above for SCMs.)\\
This approach will also cover remarks given only to transport knowledge to the code's author.\\
The remark positions are more likely to be close to the remark's trigger: The line where a review remark is anchored is probably close to the line that triggered this remark.\\
One high-level remark can lead to several changes in the code when fixed (like in the example with the typo in the method name given in Section~\ref{sec:remarksAndTriggers}). The review tool stores these conceptual remarks, whereas the SCM stores the low-level changes.\\
The number of remarks is well-defined, in contrast to the data from the SCM where it is unclear whether two consecutive changed lines should be regarded as one or two remarks (or possibly even more remarks).\\
Remarks for which the fixing was postponed and not done in the respective ticket can be extracted (also in contrast to the SCM data).\\ \hline
\end{tabularx}
\end{table}

There are two possibilities to gather review remarks from software repositories: 
(1)~Communicated review remarks can be extracted from the repository of the code review tool; or
(2)~fixed review remarks can be extracted from the SCM.
Although there usually is a high overlap between communicated and fixed remarks, there is a subset of non-fixed remarks that cannot be found in the SCM and there are remarks that the reviewer fixes ``on the fly'' without recording them in the review discussion\footnote{The review tool used in the case company allows reviewers to leave remarks for their on-the-fly fixes, but this is used only for a fraction of the fixes.}.
Table~\ref{tab:discussionDataSource} contains a detailed discussion of benefits of both approaches to extract review remarks.
There are strong reasons for both options, but we decided to use the SCM data in the current case study. One of the main reasons for this choice was that there are historically many reviews in the case company done without a tool that captures the necessary data and we wanted to include these reviews in our analysis. This decision has a significant impact on the type of noise occurring in the data. We will discuss resulting problems in Section~\ref{sec:futureWork}.

\subsection{Determinining Review Commits}

\begin{figure}[b]
\centering
  \includegraphics[width=\textwidth]{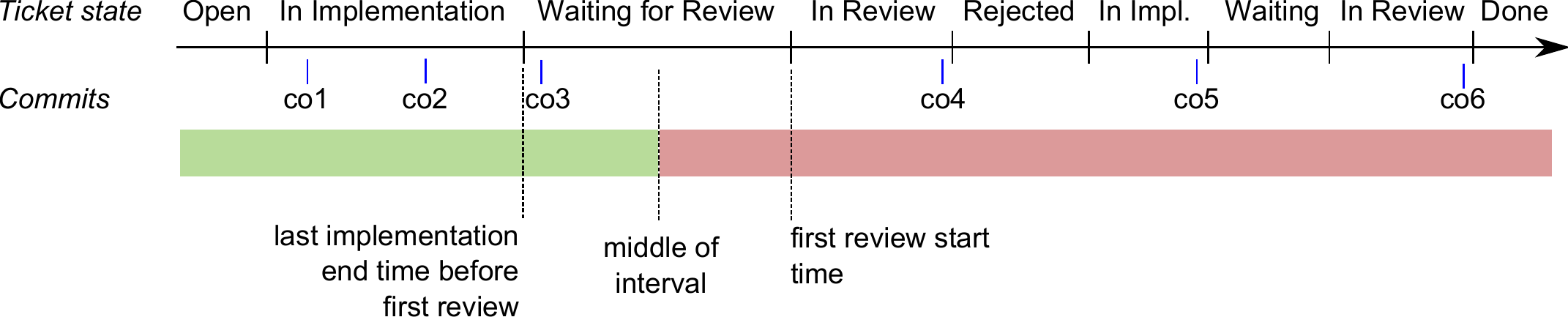}
\caption{Example of how to decide which commits for a ticket shall be regarded as ``review commits''}
\label{fig:reviewCommits}
\end{figure}

Having decided that we want to count all changes done in ``review commits'' as review remarks, the next question is how to decide whether a commit is a review commit for a certain ticket. We base our corresponding algorithm on the assumption that it is known which commits belong to a certain ticket and that a commit belongs to only one ticket. The case company has a hook script in the SCM that demands a ticket ID at the start of every commit message, which makes this information easy to extract. Figure~\ref{fig:reviewCommits} shows how the classification of the commits is done: During its lifetime, a ticket can have one of several states, of which ``in implementation'' and ``in review'' are the most important here. ``In implementation'' means that the code's author is currently working on the ticket, whereas ``in review'' means that a reviewer is checking the code. As we want to include changes performed as a reaction to review remarks, as well as changes performed by the reviewer when fixing on-the-fly, every commit after the start of the first review can be regarded as a ``review commit''. Complications arise because (1)~a ticket's state is sometimes changed before the author commits the respective changes (\eg ``co3'' in Figure~\ref{fig:reviewCommits}), and because (2)~in non-tool reviews, developers sometimes forget to change the ticket state when starting their work and do so later. To deal with these complications, we heuristically use the middle of the interval between the end of the last pre-review implementation phase and the start of the first review as the split point.

\subsection{Finding Potential Triggers}
\label{sec:tracingAlgorithm}

Given a set of review commits, the changes (\ie remarks) in these commits have to be traced back to their potential triggers. For now, we will use the finest possible granularity of tracing every changed line, as the data can be aggregated later if needed.

As mentioned above, our tracing algorithm is based on the intuition that a remark's trigger is usually close to the position of the remark. In the simplest case, some line was added or changed during implementation, and some defect in that line was found and fixed in the review phase. This is quite similar to the SZZ algorithm~\citep{Sliwerski2005} that is often used in defect prediction studies. But there's a difference to SZZ: In our case, we know that a trigger must be one of the changes in the ticket's implementation commits\footnote{``Trigger'' in this case does not necessarily mean that the root cause for the remark (\eg a defect) was injected with an implementation commit. It just means that the remark was created because of reviewing that part of the code.}. So in contrast to SZZ, we cannot stop once we found the previous change of the remark's line (\eg via ``blame''). If that previous change is not in one of the implementation commits, we need to skip it and trace back further\footnote{Similar to how \cite{kim2006automatic} skip changes that cannot be the cause of a defect in their improved version of SZZ}. We might find no potential trigger this way, for example, if there was a change in a method during implementation, and in the review it was found that another part of the method has to be changed, too. In this case, the trigger is still close to the remark (\ie in the same method), but it is not in the same line. This means that the search scope needs to be expanded. These considerations lead to the tracing algorithm outlined in Figure~\ref{fig:traceAlgorithm}. The full details, including some performance optimizations, can be seen in the Java implementation in our online materials~\citep[TraceReviewChanges.java]{baum2018predictionOnlineMaterials}.

\begin{figure}
\begin{lstlisting}[breaklines=true,language=pseudo]
function traceTicket(ticket)
   foreach review commit c in ticket
      traceCommit(c)

function traceCommit(commit)
   foreach change part p in commit
      if p is addition of whole file
         assign whole ticket as potential trigger for p
      else if p is deletion of whole file
             or p is rename of whole file without further changes
             or p is change in binary file
             or p shall be forced to file scope (e.g. file too large)
         traceWithScopeAndExpandIfNeeded(create a scope object for the whole file in p)
      else
         traceWithScopeAndExpandIfNeeded(create a line range scope for each single line in p)

function traceWithScopeAndExpandIfNeeded(scope)
   do
      found := traceWithScope(scope)
      if found == AT_LEAST_ONE_TRIGGER_FOUND
         return
      else
         //scope expansion happens depending on the file type
         scope := expand(scope)
   while scope could be expanded
    
   //the largest (implicit) scope is "whole ticket"
   assign whole ticket as potential trigger for p

function traceWithScope(scope)
   if scope is line range scope
      prevChange := 
              last change to scope.lineRange before scope.commit
   else
      prevChange := last change to scope.file before scope.commit   
   
   if prevChange not found
      return NO_TRIGGER_FOUND
        
   if prevChange is implementation commit for ticket
      assign change parts in prevChange as potential trigger(s) for scope.remark
      traceWithScope(adjust scope to commit before prevChange)
      return AT_LEAST_ONE_TRIGGER_FOUND
   else
      return traceWithScope(adjust scope to commit before prevChange)
\end{lstlisting}
\caption{Algorithm for Tracing Review Remarks to Potential Triggers (simplified)}
\label{fig:traceAlgorithm}
\end{figure}

\begin{figure}
	\centering
		\includegraphics[width=0.60\textwidth]{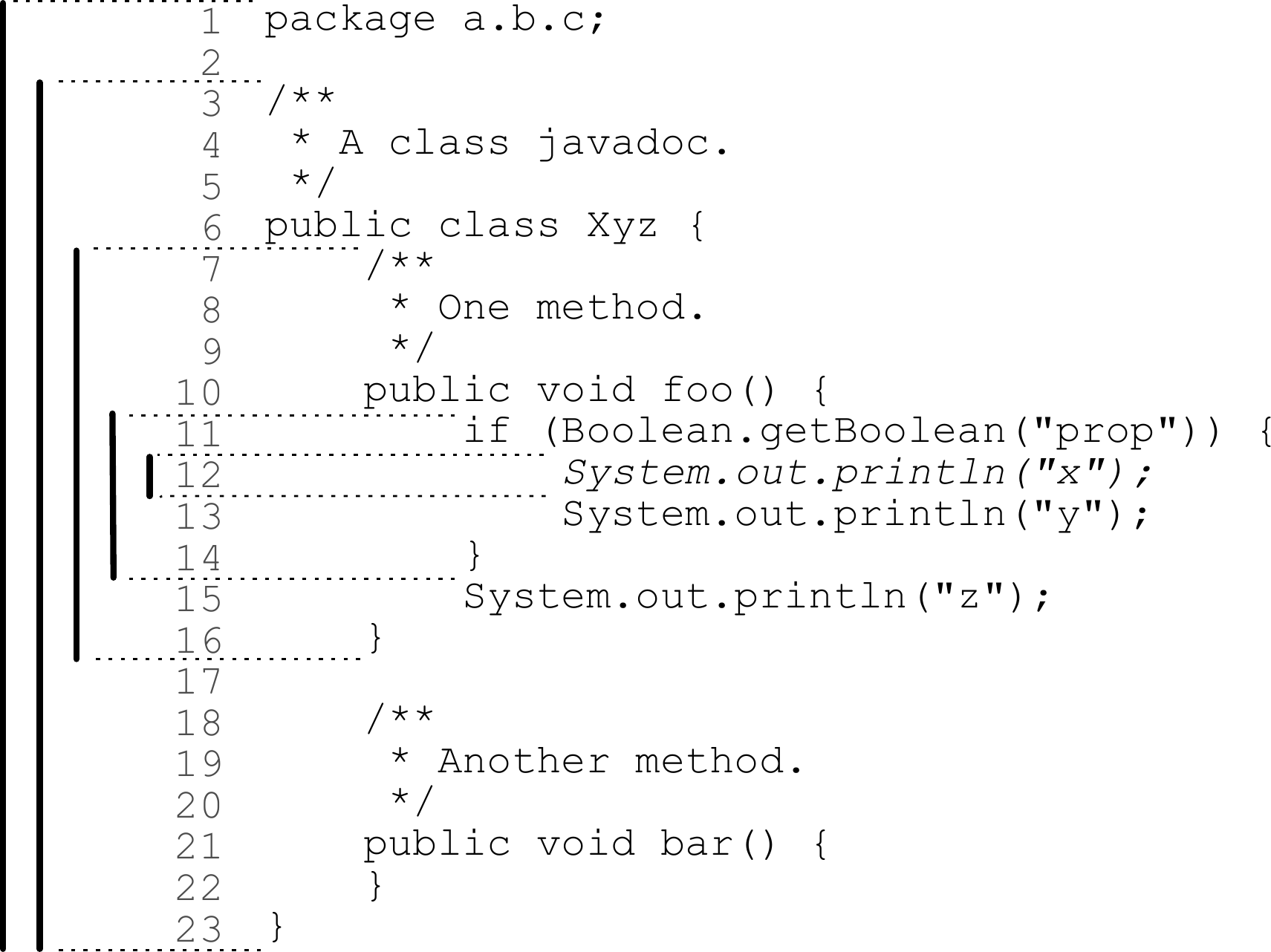}
	\caption{Example for scope expansion in a Java file, starting with the single line scope in line~12}
	\label{fig:scopeExpansionExample}
\end{figure}

As motivated from the example above, the expansion of the search scope should take the structure of the source file into account. We implemented corresponding parsers for Java and XML (and derivatives), the file types most relevant for our case study. If no potential trigger for a change is found in a single line, we first look for triggers in the line's block, then in the enclosing blocks, in the containing method, the containing class, until the whole file becomes the search scope. Figure~\ref{fig:scopeExpansionExample} shows an example of this scope expansion in a Java file. In case there was no implementation commit in the whole file, we resort to marking all changes in the whole ticket as potential triggers. For text files not supported by one of our specialized parsers, we directly move from the line scope to the file scope.

Another important detail of our algorithm is that it does not stop when the first potential trigger for a remark is found, but instead keeps on searching for further triggers with the same scope. The intuition here is that code might be added in one implementation commit, with a minor change (\eg fixing a code style issue found by static analysis) done in a later implementation commit. If the algorithm would stop after the first potential trigger, it would only find the minor change, whereas the initial addition of the code is more likely to be the ``real'' trigger.

\rqAnswer{RQ$_{2.1}$}{Review remarks can be extracted either from code review repositories or from review commits in the SCM. We focus on the latter and propose the algorithm in Figure~\ref{fig:traceAlgorithm} to associate review remarks (\ie changed lines in review commits) with potential triggers (\ie changed lines in implementation commits for the same ticket). The algorithm finds these triggers by tracing back in history, expanding the search scope if no matching trigger could be found with the smaller scope. It is based on two main assumptions: There exists at least one trigger for each review remark, and the most likely triggers are close\footnotemark{} to the remark.}
\footnotetext{with closeness defined according to a suitable scope concept for the file type}

\subsection{Comparison of Our Tracing Approach to SZZ}
\label{sec:comparisonToSzz}

The reasoning in Section~\ref{sec:classificationImproveReview} shows that review remark prediction differs from defect prediction. Nevertheless, there are similarities, and it is yet to be shown that the simple SZZ approach~\citep{Sliwerski2005} that is commonly used for tracing in defect prediction studies is not suitable to determine review remark triggers. We perform SZZ-style tracing (with ``git blame'') for all review remarks extracted from our dataset and compare the results to the results of our tracing algorithm.

\begin{figure}
	\centering
		\includegraphics[width=0.80\textwidth]{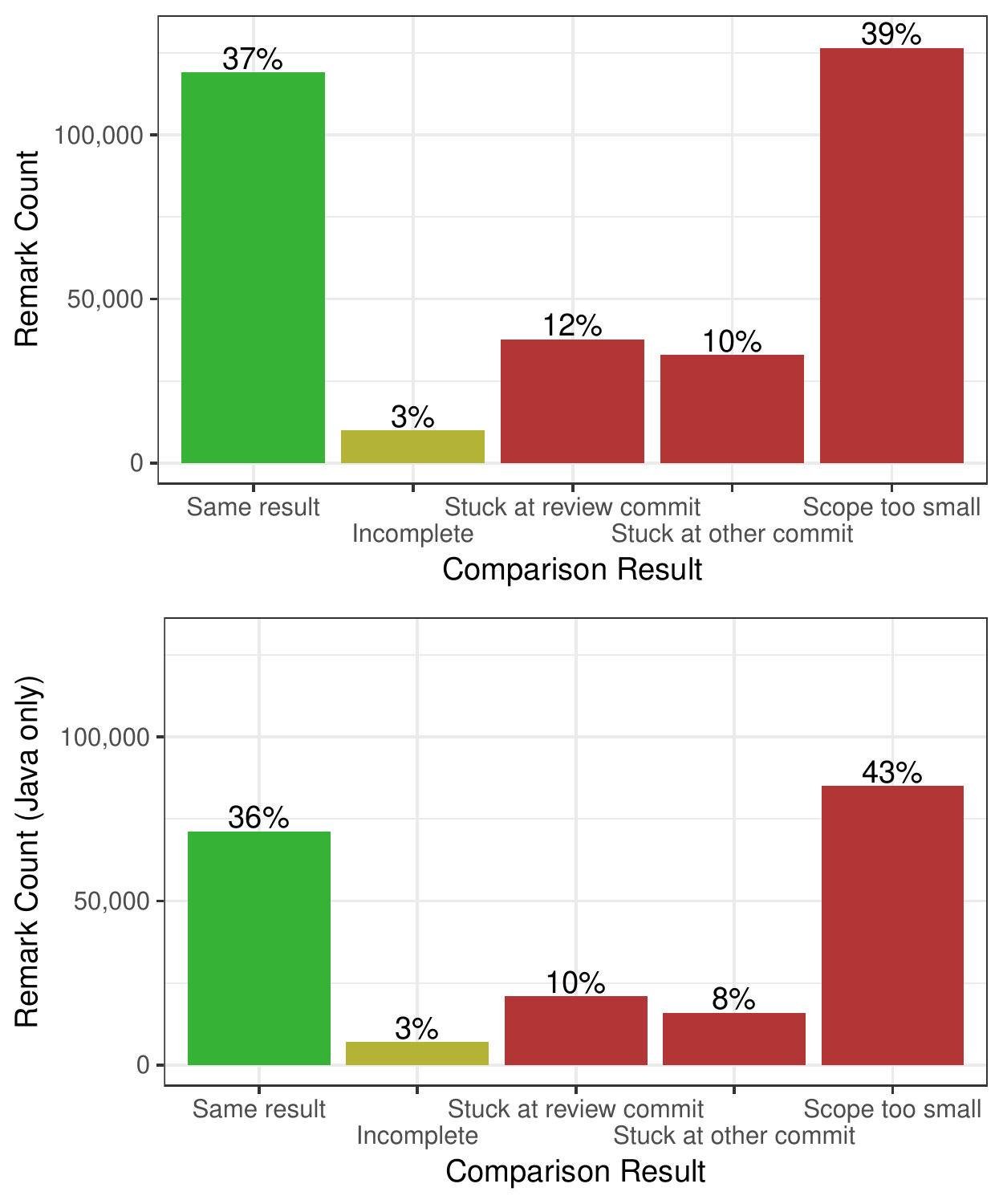}
	\caption{Comparison of the tracing approach from the current article and SZZ, both for all remarks (\ie changed lines/files in review commits) and for remarks in Java files only.}
	\label{fig:comparisonSzz}
\end{figure}

The analysis was performed on the raw data, \ie with line granularity. Figure~\ref{fig:comparisonSzz} shows the results. Of \dtaSzzTotalRemarks{} remark records, only \dtaSzzSameResultCount{} (\dtaSzzSameResultPercent{}\%) are traced with the same result for our approach and SZZ. For \dtaSzzIncompleteCount{} (\dtaSzzIncompletePercent{}\%), our approach returns further potential triggers because it does not stop at the first candidate. For \dtaSzzDiffCount{} (\dtaSzzDiffPercent{}\%) the results differ completely: For \dtaSzzStuckCount{} (\dtaSzzStuckPercent{}\%) the tracing is stopped prematurely either at a commit from another ticket or at a review commit. For \dtaSzzScopeCount{} (\dtaSzzScopePercent{}\%) SZZ does not find a trigger because it does not enlarge the search scope when no trigger is found. Figure~\ref{fig:comparisonSzz} also shows that the distribution for the subset of remark lines in Java source files is similar.

\rqAnswer{RQ$_{2.2}$}{For \dtaSzzDiffPercent{}\% of the review remarks, using SZZ to trace remarks to triggers would lead to entirely different results than our tracing algorithm. The majority of differences is caused by not enlarging the search scope when no trigger is found among the commits of the ticket under review.}

\section{Approach to Build a Prediction Model from the Mined Data}
\label{sec:howToBuildModel}

With the algorithm described in Section~\ref{sec:howToExtract} a mapping between change parts from the implementation commits and review remarks/changes can be created. We will now discuss how this data can be used to construct a model that identifies change parts to be left out from reviews.

\subsection{Intended Characteristics of the Model}
\label{sec:characteristicsOfTargetModel}

Various types of models are used in data mining techniques, \eg rules, neural networks, regression models, etc. A suitable model for our study should meet the following requirements:
\begin{itemize}
\item The constructed model must be able to reach an adequate profit, as defined in Section~\ref{sec:targetMetric}.
\item The model shall be discussed with the development team of the case study company, both to achieve better results and to increase user acceptance~\citep{tan2015online}. Therefore, it needs to be understandable by human developers.
\item For each review, it shall be transparent why certain parts of the change are left out from the review, \ie its decisions shall be ``explainable''\citep{dam2018explainable}. Furthermore, the reviewer shall be able to override this decision for certain parts.
\end{itemize}

To cross-check the requirements, we performed a survey among all available software developers of the partner company. The survey was handed out on paper and collected anonymously in the following days. The questionnaire asked for a rating of the importance of several requirements for the model and the modeling process on a 7-point Likert-type scale. It furthermore asked for a rating of several granularities to group code changes in the review into review remarks. For both questions, there was room for explanations and other free-text comments. We received 13 responses, of which 3 answered that they are not experienced enough with the topic to reply to the remaining questions. Our analysis is based on the remaining 10 responses. The survey form and the detailed results can be found in the online material~\citep{baum2018predictionOnlineMaterials}.

\begin{table}
	\centering
	\caption{Survey Results for the Importance of Various Requirements for the Prediction Model and Mining Process. All ratings are on a scale from 1 (not important at all) over 4 (neutral) to 7 (extremely important). The requirements are translations of the German originals. Rows are ordered by mean rating.}
	\label{tab:modelRequirementsResults}
	  \def\arraystretch{1.2}
		\begin{tabularx}{\textwidth}{Xrrrr}
				  \toprule
		& \multicolumn{4}{c}{Rating} \\ \cmidrule(r){2-5}
		Requirement & Mean & Median & Min. & Max. \\ \midrule
The model classifies as few change parts as possible as ``no review needed'' by error (\ie the leaving out leads to as few oversights as possible) &
5.9 & 7.0 & 3 & 7 \\
\rowcolor{tableLight}
The model can be evaluated quickly at review start &
5.5 & 5.5 & 3 & 7 \\
I can see why a certain change part was classified as ``no review needed'' &
5.3 & 6.0 & 1 & 7 \\
\rowcolor{tableLight}
The parts of the model have been checked by the development team before deployment to production &
5.2 & 5.0 & 4 & 7 \\
I can override/disable parts of the model so that the respective change parts remain in the review scope &
5.2 & 5.0 & 2 & 7 \\
\rowcolor{tableLight}
The initial creation of the model from the raw data requires little human interaction &
4.7 & 5.0 & 1 & 6 \\
The model classifies as many change parts as possible as ``no review needed'' &
4.4 & 4.5 & 1 & 7 \\
\rowcolor{tableLight}
The initial creation of the model from the raw data is quick &
4.0 & 4.0 & 1 & 7 \\ \bottomrule
    \end{tabularx}
\end{table}

Table~\ref{tab:modelRequirementsResults} shows the aggregated results for the requirements' importance. The most important requirements are a low number of misclassifications and a quick review start. Here, ``quick review start'' is more a restriction on the usable features of the data, as all common model types are quick to evaluate. Next, and also with high importance on average, come the three requirements on the understandability of the model. The requirements with the least relative importance are the quick and effortless creation of the model and maximization of the number of change parts that the model classifies as ``no review needed''. When we looked at the latter result in detail, we found a large spread in the answers, with some developers considering this as very important and others of the opposite opinion. One of the developers who considered a high number of ``no review'' classifications as unimportant explained his answer: ``I consider leaving out change parts in reviews as dangerous because it can give a false sense of security. Personally, I prefer thorough code reviews.''
 
For the second question in the survey, the right granularity for counting missed review remarks, the developers preferred counting change parts, followed by counting lines. Counting at a coarser granularity, \ie files or whole tickets, was considered inadequate by the majority of respondents.

\rqAnswer{RQ$_{3.1}$}{The three most important requirements from the team are that the model provokes as few missed review remarks as possible, does not lead to additional waiting time in the review, and is transparent to and checked by the team.}

Concerning this section's goal to select an adequate type of model, the survey results support the initially stated requirements.
A rule-based model seems well-suited to satisfy these requirements: Rules are a well-known concept for software developers, and a rule-based model is relatively easy to analyze manually. Our classification task is binary, classifying each change part as either ``needs no review'' or ``review''. Based on the intuition that it is possible to find some rules for change parts that definitely need to be reviewed, other parts that definitely need no review and a harder to classify rest that will conservatively be kept for review, we envision rules of the following form:
\begin{lstlisting}
skip when one of
   (... and ... and ...)
   or (...)
   ...
unless one of
   (... and ... and ...)
   or (...)
   ...
\end{lstlisting}
Or, put mathematically: $skip := \bigvee_{r \in incl}\left(\bigwedge_{c \in r} c \right) \wedge \neg \bigvee_{r \in excl}\left(\bigwedge_{c \in r} c \right)$. This notion allows certain rules to be written more concisely than a simple disjunctive normal form but is still quite simple and easily explainable. Furthermore, every single rule in the disjunctions can be treated as a separate nugget of knowledge and can be used to explain and override the decisions of the review tool. We restrict ourselves to propositional logic for the single conditions: ``less or equal'' and ``greater or equal'' for numeric features and ``equals'' and ``not equal'' for categorical features.

\subsection{Mining Rules from the Extracted Data}
\label{sec:miningAlgorithm}

There are two complications when trying to use standard data mining algorithms on the extracted data: (1)~The standard algorithms optimize for accuracy, not for profit. (2)~They cannot take into account that remarks are not just simple classification labels but are objects with a distinguishable identity: A mining algorithm can exploit the identity of the remarks because it is sufficient to cover only one of the potential triggers for each remark. As an example, consider two potential triggers, one an addition of complex code and the other a minor change. Without taking into account that only one of these triggers is needed, a rule of the form ``minor changes need no review'' would be unnecessarily seen as inaccurate. Section~\ref{sec:results} will show that the generic rule mining algorithm RIPPER~\citep{Cohen1995} indeed creates inferior results in our context.

One of the variants of profit, as introduced in Section~\ref{sec:targetMetric}, is our primary target metric. But simply maximizing it on the training set can lead to overfitting. Simpler rules and rules that have fewer features can work better on unseen data, even though they seem worse on the training set. We believe that this problem is best dealt with by letting the development team select the final ruleset from a number of candidates with different characteristics. Simpler rules have the additional benefit of being easier to explain, and rules with fewer features lead to less implementation effort in a tool. Therefore, we treat the mining task as a multi-objective problem: The basic objectives are ``maximize saved effort'', ``minimize the number of missed review remarks'', ``minimize rule complexity'' and ``minimize the number of used features''. Saved effort and missed review remarks are used in favor of profit because the latter can be derived from the former and do not need an estimate of the cost of missed remarks. More details on the finally used objective vector will be given in Section~\ref{sec:iterativeImprovement}.

\begin{figure}
	\centering
		\includegraphics[width=0.55\textwidth]{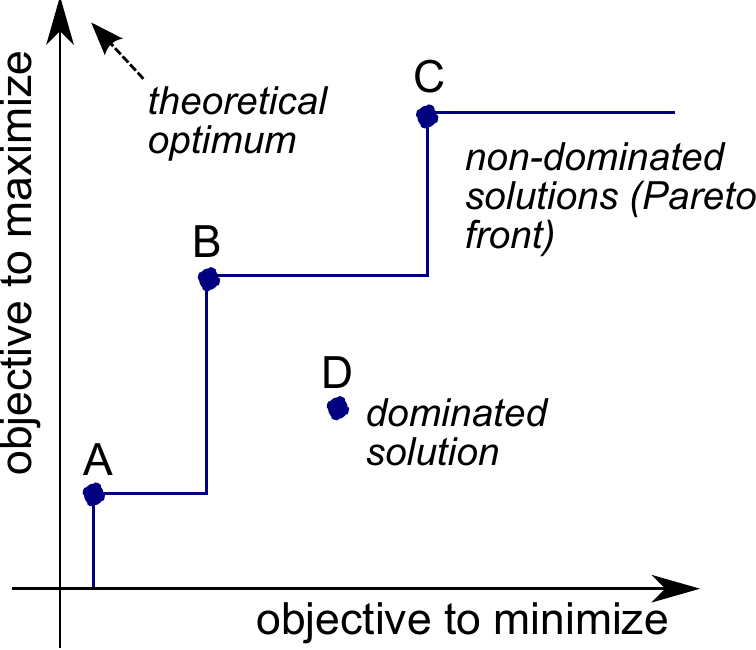}
	\caption{Example to illustrate the concept of Pareto-optimality. Solution B dominates solution D because it is better in both objectives. The other solutions do not dominate each other because they are better for one but worse for the other objective. They form the Pareto front.}
	\label{fig:paretoExample}
\end{figure}

To be able to take the above problems and the multi-objective nature of the problem into account, we use a heuristic to induce the rules. The heuristic is based on the GRASP-PR meta-heuristic~\citep{resende2016optimization}, with the following key steps:
\begin{enumerate}
\item Use a randomized greedy heuristic\footnote{In contrast to a usual deterministic greedy heuristic, a randomized greedy heuristic might divert from the best greedy choice based on a randomized condition.} for set cover to select a subset of the potential triggers as the change parts to cover by the rules. After this step, every change part is classified as either NO\_TRIGGER or TRIGGER.
\item Use a randomized greedy separate-and-conquer heuristic to derive a ruleset candidate based on the dataset created in the previous step.
\item Optimize the initial ruleset by local search, \eg by leaving out conditions or adjusting the split point for numeric conditions. All rulesets generated during the local search are added to an archive of Pareto-optimal solutions (Figure~\ref{fig:paretoExample} illustrates Pareto-optimality).
\item Select one of the existing and one of the new rulesets to perform ``path relinking'': One ruleset is gradually transformed into the other, and the rulesets created in this way are again added to the archive of Pareto-optimal results
\end{enumerate}

These steps are iterated continuously in the background, while the user can explore the data and the results found so far. Based on insights gained this way, the user can provide feedback to guide future iterations of the heuristic. In this way, we perform data mining with the human in-the-loop \citep{ankerst2002report}. To keep the current article focused, we cannot describe all the details of the used rule mining system here. A detailed description is available in a separate report~\citep{baum2018miningTool}.

\section{Feature Selection}
\label{sec:features}

A data mining model can only deliver good results if the used features fit the task. To select the features for the current study, we used two sources: (1)~Features that showed good results in defect prediction studies, based on a survey of the literature. (2)~Iterative analysis and idea generation (``open coding'') based on the code change data. By combining these sources, we account for the similarity between defect prediction and review remark prediction but also for the likely differences and more specialized features. We had to restrict the selection to a subset of the potentially applicable features to limit the effort for this part of the study.

To identify features from the literature, we looked for defect prediction studies and especially studies that compare the relative suitability of features for defect prediction. We identified these studies using searches on Google Scholar and performed snowballing~\citep{Wohlin2014} for further studies, but did not perform a formal systematic literature review. Few studies use change part granularity for prediction, so the features were partly adapted to fit our context. The studies selected for analysis are: \citep{menzies2010defect,Radjenovic2013,Giger2012,Nagappan2005,Shihab2012,Ratzinger2007,mockus2000predicting,mcintosh2017fix,d2010extensive,arisholm2010systematic,soltanifar2016predicting,chen2017applying,shivaji2013reducing,meneely2008predicting,shin2009does,eyolfson2011time,shihab2012exploration,osman2017empirically}.

As second feature source, we used an inductive approach, inspired by qualitative approaches for hypothesis generation~\citep{glaser1967discovery}: Initially, we sampled some code changes from the data and assigned ``codes'' to the change parts, similar to ``open coding'' in qualitative data analysis. Later, we extended the data mining tool with a feature to sample misclassified change parts. These were again analyzed for missing features that could explain the misclassification.

Tables \ref{tab:features1}~to~\ref{tab:features4} show the final selection of features, with a brief explanation and the source for each feature. `Source' means the source responsible for the inclusion in this study and does not contain an exhaustive list of all studies that use that feature.

\renewcommand{\floatpagefraction}{0.05}

\begin{table}[p]
	\centering
	\caption{Final Selection of Change Part Features used as Input for the Mining (1/4)}
	\label{tab:features1}
		\begin{tabular}{>{\raggedright}p{2.4cm}|>{\raggedright}p{5.6cm}|>{\raggedright\arraybackslash}p{2.5cm}}
			\rowcolor{tableDark} Name and Type & Description & Source / Inspired by \\ \hline \hline
			
			\rowcolor{tableMedium}\multicolumn{3}{l}{Ticket and commit granularity:} \\ \hline
issueType \emph{(nominal)} & Type of the Jira ticket (\eg bug or user story task) & \citet{Shihab2012} \\ \hline
author \emph{(nominal)} & User ID for the author of the commit & inductive \\ \hline
authorDay \emph{(nominal)} & Weekday of the commit & \multirow{2}{2.5cm}{\citet{eyolfson2011time,Shihab2012}} \\ \cline{1-2}
shiftedAuthorHour \emph{(numeric)} & Hour of the time of the commit. The value is shifted so that 0 stands for 6 AM. In this way, ``night'' vs ``day'' can be expressed with a single comparison. &  \\ \hline
fileCountInCommit \emph{(numeric)} & Count of files that were changed in the commit. & inductive \\ \hline
hunkCountIn\-Commit \emph{(numeric)} & Count of change parts in the commit (changes in binary files count as 1) & inductive \\ \hline
commitContains\-Test \emph{(boolean)} & ``true'' iff the commit contains changes to test code & \citet{Ratzinger2007} \\ \hline
			
			\rowcolor{tableMedium}\multicolumn{3}{l}{File granularity:} \\ \hline
binary \emph{(boolean)} & ``true'' iff the file is treated as binary. Very large text files ($\geq$ 1 MiB) are also treated as binary. & inductive \\ \hline
filetype \emph{(nominal)} & Extension of the filename (\eg ``java'' or ``txt'') & inductive \\ \hline
srcdir \emph{(nominal)} & Classification of the file in the project: ``src'' (production code), ``test'' (test code), ``testdata'' or ``resources'' & inductive \\ \hline
project \emph{(nominal)} & Project to which the file belongs. In the case company, there are also some pseudo projects, \eg for common build scripts or external dependencies. & inductive \\ \hline
frequentFilename \emph{(nominal)} & Filename (without path, but with extension), but set only when it is one of the 20 most common filenames. In the case company there are some very common filenames that denote specific roles, \eg ``Messages.java'' or ``Logger.java''. & inductive \\ \hline
fileAgeDays \emph{(numeric)} & Number of days since the creation (initial commit) of the file. & \citet{shin2009does,Radjenovic2013,osman2017empirically} \\ \hline
fileCommitCount \emph{(numeric)} & Number of commits to the file since its creation. & \citet{meneely2008predicting,arisholm2010systematic,Shihab2012,mcintosh2017fix} \\ \hline
		\end{tabular}
\end{table}

\begin{table}[p]
	\centering
	\caption{Final Selection of Change Part Features used as Input for the Mining (2/4)}
	\label{tab:features2}
		\begin{tabular}{>{\raggedright}p{2.4cm}|>{\raggedright}p{5.6cm}|>{\raggedright\arraybackslash}p{2.5cm}}
			\rowcolor{tableDark} Name and Type & Description & Source / Inspired by \\ \hline \hline
			
			\rowcolor{tableMedium}\multicolumn{3}{l}{File granularity (continued):} \\ \hline
distinctFileAuthor\-Count \emph{(numeric)} & Number of distinct authors of the file since its creation. & \citet{meneely2008predicting,shin2009does,arisholm2010systematic,Shihab2012,Giger2012,osman2017empirically,mcintosh2017fix} \\ \hline
newLineCountIn\-File \emph{(numeric)} & Total number of lines in the file (after the commit) & \citet{menzies2010defect,shihab2012exploration} \\ \hline
recentProject\-Ownership \emph{(numeric)} & Ratio of the number of commits to the file's project in the last year by the author to the number of commits to the file's project in the last year by all authors. & \citet{mockus2000predicting,soltanifar2016predicting} \\ \hline
commitsSinceLast\-RemarkForAuthor\-InProject \emph{(numeric)} & Number of commits since the file's author last received a review remark in the file's project.\vspace{1em} & \multirow{2}{2.5cm}{\citet{shin2009does,d2010extensive,Shihab2012,Radjenovic2013,soltanifar2016predicting,osman2017empirically}} \\ \cline{1-2}
commitsSinceLast\-RemarkInFile \emph{(numeric)} & Number of commits since the file last received a review remark.\vspace{1.1cm} &  \\ \hline
hunkCountInFile \emph{(numeric)} & Count of change parts in the file & inductive \\ \hline
changetype \emph{(nominal)} & Git's classification of the change to the file (MODIFY/ADD/RENAME/\allowbreak{}DELETE/COPY). & inductive \\ \hline
gitSimilarity \emph{(numeric)} & Git's similarity statistic for the file content (\ie 100 when the content stayed the same) & inductive \\ \hline
newShareOfLines\-InFile \emph{(numeric)} & Ratio of the changed lines to the total number of lines (in the new file version). & \citet{Ratzinger2007} \\ \hline
isNodeModules \emph{(boolean)} & ``true'' if the file path denotes a commited external dependency from npm (node.js package manager) & inductive \\ \hline

			\rowcolor{tableMedium}\multicolumn{3}{l}{Change part granularity:} \\ \hline
oldHunkSize \emph{(numeric)} & Number of lines of the change part in the old file version & inductive \\ \hline
newHunkSize \emph{(numeric)} & Number of lines of the change part in the new file version & inductive \\ \hline
changeInHunkSize \emph{(numeric)} & newHunkSize minus oldHunkSize & inductive \\ \hline
		\end{tabular}
\end{table}

\begin{table}[p]
	\centering
	\caption{Final Selection of Change Part Features used as Input for the Mining (3/4)}
	\label{tab:features3}
		\begin{tabular}{>{\raggedright}p{2.4cm}|>{\raggedright}p{5.6cm}|>{\raggedright\arraybackslash}p{2.5cm}}
			\rowcolor{tableDark} Name and Type & Description & Source / Inspired by \\ \hline \hline
			
			\rowcolor{tableMedium}\multicolumn{3}{l}{Change part granularity (continued):} \\ \hline
commentLine\-CountOld \emph{(numeric)} & Number of comment lines in the old side of the change part & \multirow{3}{2.5cm}{\citet{menzies2010defect,Giger2012}} \\ \cline{1-2}
commentLine\-CountNew \emph{(numeric)} & Number of comment lines in the new side of the change part &  \\ \cline{1-2}
changeInComment\-LineCount \emph{(numeric)} & commentLineCountNew minus commentLineCountOld &  \\ \hline
oldBlockCount \emph{(numeric)} & Number of Java blocks (\ie braces pairs) in the old side of the change part. & \multirow{3}{2.5cm}{\citet{menzies2010defect}} \\ \cline{1-2}
newBlockCount \emph{(numeric)} & Number of Java blocks (\ie braces pairs) in the new side of the change part. &  \\ \cline{1-2}
changeInBlock\-Count \emph{(numeric)} & newBlockCount minus oldBlockCount &  \\ \hline
responseForHunk\-Old \emph{(numeric)} & RFC metric (Response For a Class; approx. number of distinct method calls) restricted to the code in the old side of the change part & \multirow{2}{2.5cm}{\citet{d2010extensive,arisholm2010systematic,Giger2012,osman2017empirically,chen2017applying}} \\ \cline{1-2}
responseForHunk\-New \emph{(numeric)} & RFC metric (Response For a Class; approx. number of distinct method calls) restricted to the code in the new side of the change part &  \\ \hline
changeInResponse\-ForHunk \emph{(numeric)} & responseForHunkNew minus responseForHunkOld & \citet{arisholm2010systematic,d2010extensive} \\ \hline
whitespaceOnly \emph{(boolean)} & ``true'' iff only whitespace was changed for the change part. & inductive \\ \hline
packageAndImport\-Only \emph{(boolean)} & ``true'' iff only package declarations and import statements were changed for the change part. & inductive \\ \hline
finalChangeOnly \emph{(boolean)} & ``true'' iff the change consists only of adding or removing Java's ``final'' keyword. & inductive \\ \hline
nonnlsChangeOnly \emph{(boolean)} & ``true'' iff the change consists only of changes to the marker comments and annotations for non-internationalized string constants. & inductive \\ \hline
visibilityChange\-Only \emph{(boolean)} & ``true'' iff only visibility modifiers (``public'', ``private'', \dots) were changed. & inductive \\ \hline
overrideAnnotation \emph{(nominal)} & Denotes which side of the change part contains an ``@Override'' annotation (none/old/new/both) & inductive \\ \hline
		\end{tabular}
\end{table}

\clearpage

\renewcommand{\floatpagefraction}{0.5}

\begin{table}[t]
	\centering
	\caption{Final Selection of Change Part Features used as Input for the Mining (4/4)}
	\label{tab:features4}
		\begin{tabular}{>{\raggedright}p{2.4cm}|>{\raggedright}p{5.6cm}|>{\raggedright\arraybackslash}p{2.5cm}}
			\rowcolor{tableDark} Name and Type & Description & Source / Inspired by \\ \hline \hline
			
			\rowcolor{tableMedium}\multicolumn{3}{l}{Change part granularity (continued):} \\ \hline
entropyCbMax, entropyCbUppQuar, entropyCbMed, entropyCbSum, entropyCbAvg \emph{(numeric)} & This group of features conceptually denotes the ``surprisingness'' of the new code given the old codebase. Formally, we use the SLP library by \citet{hellendoorn2017deep} to compute an entropy for each token on the new side of the change part, with smaller values for less surprising tokens. These per token entropies are then combined to find the maximum (``Max''), 75\% quantile (``UppQuar''), median (``Med''), sum (``UppQuar'') and mean (``Avg'') for the change part. & inductive, \citet{hellendoorn2017deep} \\ \hline
entropyReMax, entropyReUppQuar, entropyReMed, entropyReSum, entropyReAvg \emph{(numeric)} & This group of features conceptually denotes the ``surprisingness'' of a change part given the earlier change parts under review. Like the ``entropyCb\dots'' group of features, it uses the SLP library by \citet{hellendoorn2017deep} and combines the per token values differently for each feature (``Max'', ``UppQuar'', \dots). Pre-processing is performed to make the library work on code changes instead of code; its details can be seen in the online material~\citep{baum2018predictionOnlineMaterials}. & inductive, \citet{hellendoorn2017deep} \\ \hline
		\end{tabular}
\end{table}

\section{Application of the Approach within the Case Study Company}
\label{sec:results}

After having outlined our approach in the previous sections, this section describes the results of applying it in the case study company.

\subsection{Extracted Data}

\begin{table}
	\centering
	\caption{Number of Commits, Change Part Records and Review Remarks in Total and per Ticket for the Extracted Training Data. Commits are subdivided into implementation and review commits. There are \dtaTicketCount{} tickets in total. All counts are after cleaning.}
	\label{tab:extractedDataStatistics}
		\begin{tabular}{lrrrrrr}
\toprule
 &
 &
\multicolumn{5}{c}{Per Ticket} \\ \cmidrule(r){3-7}

 &
 Total &
Min. &
25\% Qu. &
Median &
75\% Qu. &
Max. \\ \midrule

Commits (impl. + review) &
\dtaAllCommitsTotal{} &
\dtaAllCommitsPerTicketMin{} &
\dtaAllCommitsPerTicketLowQuart{} &
\dtaAllCommitsPerTicketMedian{} &
\dtaAllCommitsPerTicketUppQuart{} &
\dtaAllCommitsPerTicketMax{} \\

Commits (impl.) &
\dtaImplCommitsTotal{} &
\dtaImplCommitsPerTicketMin{} &
\dtaImplCommitsPerTicketLowQuart{} &
\dtaImplCommitsPerTicketMedian{} &
\dtaImplCommitsPerTicketUppQuart{} &
\dtaImplCommitsPerTicketMax{} \\

Commits (review) &
\dtaReviewCommitsTotal{} &
\dtaReviewCommitsPerTicketMin{} &
\dtaReviewCommitsPerTicketLowQuart{} &
\dtaReviewCommitsPerTicketMedian{} &
\dtaReviewCommitsPerTicketUppQuart{} &
\dtaReviewCommitsPerTicketMax{} \\

Change part records &
\dtaRecordsTotal{} &
\dtaRecordsPerTicketMin{} &
\dtaRecordsPerTicketLowQuart{} &
\dtaRecordsPerTicketMedian{} &
\dtaRecordsPerTicketUppQuart{} &
\dtaRecordsPerTicketMax{} \\

Change part rec. (Java) &
\dtaJavaRecordsTotal{} &
\dtaJavaRecordsPerTicketMin{} &
\dtaJavaRecordsPerTicketLowQuart{} &
\dtaJavaRecordsPerTicketMedian{} &
\dtaJavaRecordsPerTicketUppQuart{} &
\dtaJavaRecordsPerTicketMax{} \\

Review remarks &
\dtaRemarksTotal{} &
\dtaRemarksPerTicketMin{} &
\dtaRemarksPerTicketLowQuart{} &
\dtaRemarksPerTicketMedian{} &
\dtaRemarksPerTicketUppQuart{} &
\dtaRemarksPerTicketMax{} \\ \bottomrule
\end{tabular}
\end{table}

Before delving into the details of the mining, we give some characteristics of the extracted data. Table~\ref{tab:extractedDataStatistics} shows statistics for the extracted training data. It can be seen that the distributions per ticket are all heavily skewed, with a few huge changes. Java is the primary implementation language, and therefore the majority of changes is performed in Java files. Next in frequency of occurrence are XML schema files. XML schema is heavily used for interface definitions in the company. Most other frequently occurring changes are due to test data and committed dependencies.
The earliest analyzed commit in the training data is from March 2013, the most recent from July 2018.
Of the \dtaRecordsTotal{} change part records, \dtaNoTriggerRecordCount{} have no association to a review remark at all. Of the remaining, a minority of \dtaMustRecordCount{} records is the only trigger for at least one review remark. These records must be triggers in our approach, whereas the remaining records are only candidates for triggers.

The size of the repositories and the extracted data led to long run-times for several of the mining steps.
Extraction of the feature data was most time-consuming for the codebase entropy features, for which it took more than a week. The other extraction programs and the tracing usually completed within a few hours or days.
One assumption underlying our mining approach is that it is not a problem to let the mining run for a longer time, as long as feedback on intermediate results can be given at any time. Consequently, the mining agents ran in the background for several days. Feedback could be given as soon as the data was loaded and first rules were created, which was after less than half an hour. Our comparison with the standard rule mining algorithm RIPPER~\citep{Cohen1995} required about two hours of computation time for each ruleset.

\subsection{Iterative Improvement of the Approach}
\label{sec:iterativeImprovement}

We used the interactive rule mining system outlined in Section~\ref{sec:miningAlgorithm} to infer rules from the data. Input from domain experts on intermediate results was given by the first author and by three other developers from the industrial partner. The field notes and interaction logs can be found in the online material~\citep{baum2018predictionOnlineMaterials}. The feedback can be categorized into specific feedback on the mining results and meta-feedback on the approach. In this section, we describe the meta-feedback and the resulting improvements to the approach.

The other developers found the system helpful and considered it interesting to analyze the data with it.
Apart from minor technical problems with the rule mining system, the domain experts identified two major problems:
\begin{description}
\item[\emph{Too much focus on a small fraction of large tickets.}] Initially, the algorithm focused too narrowly on parts of the data with large-scale changes, \eg single tickets that changed many files. These are correct findings that shouldn't simply be removed as ``outliers'', but they are not very interesting because they account only for a small fraction of the tickets and the review effort.
\item[\emph{Too much noise in the data.}] Noise appears in the data in several ways: Identified remarks might not be remarks at all, or they might be traced to wrong triggers. These problems can be due to remarks that are follow-ups of other remarks (\eg rename refactorings or overrides) or remarks for code not belonging to the ticket. Another form of noise is caused by not taking the severity of the remarks into account, \eg if a whitespace correction influences the algorithm as much as a critical defect. We analyzed a random sample of 100 remarks from an early version of the data to assess the problem. When checking for quality of the remarks, we found that 19\% were no true remarks. When checking the tracing, we found that for 71\% of the remarks the tracing was OK. Of the remaining 29\%, 9\% were definitely wrong. Details are again available in the online material~\citep{baum2018predictionOnlineMaterials}.
\end{description}

\noindent Based on this feedback, we changed our rule mining approach to mitigate the problems:
\begin{description}
\item[\emph{Additional objectives.}] We initially started with four objectives: missed remark count (to minimize), saved record count (to maximize), rule complexity (to minimize) and number of used features (to minimize). These are generic objectives that could be applied to almost every data mining problem\footnote{taking into account that ``missed remark count'' is similar to ``false positives'' and ``saved record count'' is similar to ``true positives + false positives''}. Using only these objectives gives rise to the above-mentioned ``too much focus'' problem. To counter this problem, we added three further objectives: (1)~``Saved lines of code in Java files'' captures the intuition that the main effort of a review is mostly spent on the source code. (2)~``Log-transformed missed remarks'' is calculated by counting each missed remark with the logarithm of the total number of remarks in its ticket (instead of 1 as for ``missed remark count''). This objective puts less weight on large review changes, based on the intuition that these are often systematic. (3)~``Trimmed mean of saved records per ticket'' is calculated as the 20\% trimmed mean~\citep{wilcox2011introduction} of the number of saved records per ticket. It puts less weight on rare tickets that have a large number of records. As a per ticket measure, it is also easy to interpret by domain experts.
\item[\emph{Additional target functions.}] Based on the additional objectives described above, we also defined additional target functions. The most important of these is a family of per ticket cost functions. With cost defined as negative profit as given in Equation~\ref{eqn:profit} and the cost factor $c$ as introduced in Section~\ref{sec:targetMetric}, we have:
\[\mbox{cost}_{c} = \frac{r}{t}c - s\]
In this formula, $r$ is the log-transformed missed remark objective value, $t$ is the number of analyzed tickets, and $s$ is the trimmed mean of saved records per ticket.
\item[\emph{Merging of remarks with identical content.}] One of the problems with determining the review remarks based on the changes in review commits is that systematic changes lead to a large number of ``remarks'' whose number does not adequately represent the cost of missing one of them. In addition to the ``log-transformed missed remarks'' objective (see above), we merged all change parts that represent the same textual change in the same commit into one remark.
\item[\emph{Removal of certain remark types.}] Another change to get rid of remarks with minor relevance was to delete certain types of remarks: Remarks with whitespace-only changes, remarks with changes to package declaration or import statements only, and other derived changes.
\item[\emph{Manual cleaning of the data.}] We systematically checked tickets with a large number of changes or a large number of review commits and removed problematic ones, \eg  when there was a violation of the development process that led to unusable data.
\end{description}

\subsection{Rule Mining Results}
\label{sec:ruleMiningResults}

The current study is the first to use the interactive multi-objective rule mining algorithm outlined in Section~\ref{sec:miningAlgorithm}. Therefore, we compare the results from that novel approach (called ``MO\_I'' in the following) to three other approaches: The results from our multi-objective algorithm, but without human interaction (``MO\_A''), a rule set created by the RIPPER rule mining algorithm~\citep{Cohen1995}, and a baseline of skipping a randomly sampled share of the records for each ticket. To make the information on review remarks and their potential triggers (see Section~\ref{sec:remarksAndTriggers}) amenable for RIPPER, we took a conservative approach: Every record that is linked to at least one remark is labeled as ``must review'' and records without a link to a remark are labeled as ``no trigger''. A second variant of RIPPER was used in the case study, but due to a defect in our software implementation, its results had to be discarded. Nevertheless, the respective details are contained in the online material~\citep{baum2018predictionOnlineMaterials}.
 
There are no established guidelines on how to assess the mining results in a case like ours. Relying on iterative human feedback makes cross-validation impossible, and many metrics (like precision, recall or cost) are only defined for single rulesets and not for a Pareto set of rulesets. Therefore, we combine several evaluation approaches: (1)~We check where the results from RIPPER lie relative to the Pareto front of MO\_I. (2)~We select rulesets from the Pareto fronts for MO\_I and MO\_A and compare the objective values of all four rulesets. (3)~We ask the company's developers for their opinion on the rules. The current subsection presents the results for (1)~and~(2) for the training data, and Section~\ref{sec:developersOpinion} presents the results of the developer assessment and the discussion with the development team. Section~\ref{sec:performanceUnseenData} shows the results for unseen test data. For the discussion with the development team, the ruleset MO\_I was further adapted. Statistics for this adapted version (``SESSION'') will also be given in the following.

\begin{figure}
	\centering
		\includegraphics[height=0.92\textheight]{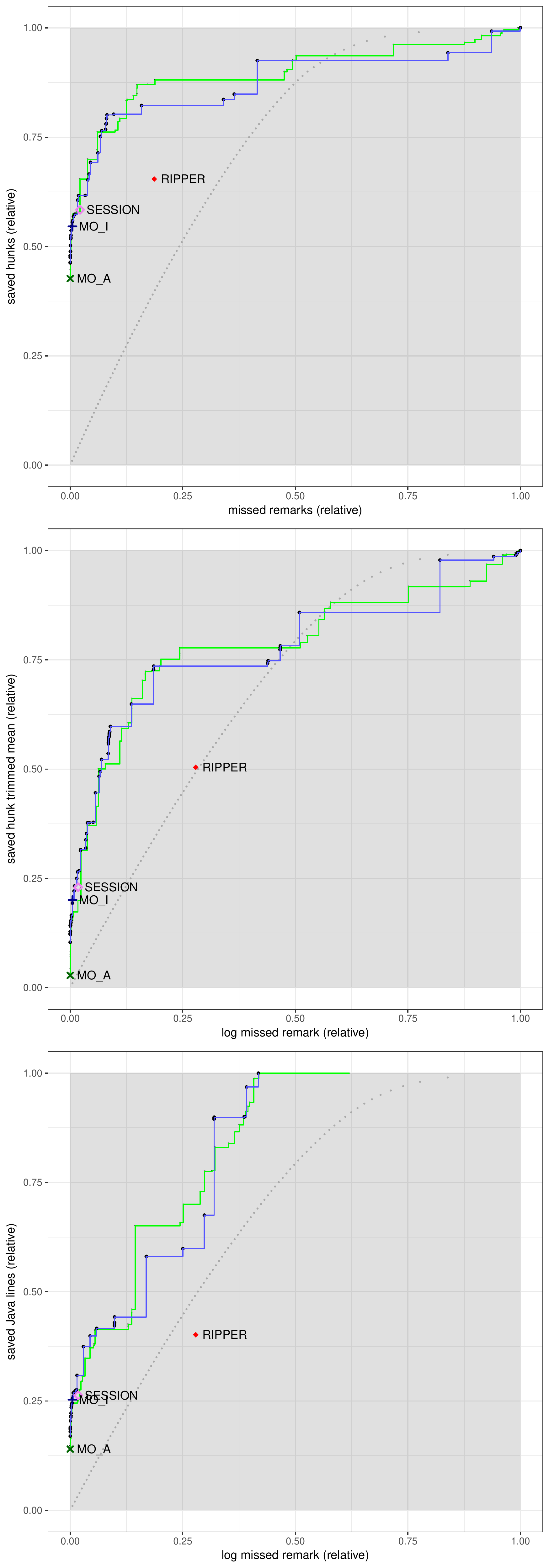}
	\caption{Pareto fronts and selected rulesets, evaluated on the training data. The plots show two-dimensional projections from the seven-dimensional objective space. The gray dots show the baseline performance of leaving out a certain percentage of records per ticket; each dot corresponds to a percentage value, with results averaged over 100 random seeds.}
	\label{fig:resultsTrainingSet}
\end{figure}

Figure~\ref{fig:resultsTrainingSet} shows projections of the Pareto front obtained by the multi-objective algorithm with and without input from domain experts. It also shows the position of the four selected rules in the objective space. RIPPER is dominated by both Pareto fronts by a large degree, and does not perform much better, and sometimes worse, than just skipping the review of random change parts. RIPPER differs from all other rulesets by having the form ``skip all except \dots''. RIPPER was insensitive to the (high) cost of missed remarks and created rulesets with many missed review remarks. The RIPPER rulesets are also more complex than the MO rulesets. The exact numbers for these and the other objectives are shown in Table~\ref{tab:objectiveValuesForSelection}.

\begin{table}
	\centering
	\caption{Objective Values for the Selected Rulesets on the Training Data.}
	\label{tab:objectiveValuesForSelection}
	\begin{threeparttable}
		\begin{tabularx}{\textwidth}{Xrrrrrrr}
\toprule
 &
\multicolumn{4}{c}{Objectives to Minimize} &
\multicolumn{3}{c}{Objectives to Maximize} \\ \cmidrule(r){2-5} \cmidrule(r){6-8}

Ruleset &
\parbox{1.0cm}{\raggedleft Compl\-exity} &
\parbox{1.0cm}{\raggedleft Feature\\Count} &
\parbox{1.1cm}{\raggedleft Missed\\Remark\\Count} &
\parbox{1.0cm}{\raggedleft Log-Transf.\\Missed\\Remarks} &
\parbox{1.0cm}{\raggedleft Saved\\Record\\Count} &
\parbox{1.0cm}{\raggedleft Tr.M.\tnote{1}\\Saved\\Records Per\\Ticket\\} &
\parbox{1.0cm}{\raggedleft Saved\\LOC\\in Java\\Files} \\ \midrule

SESSION &
\dtaSesComplexity &
\dtaSesFeatureCount &
\dtaSesMissedRemarks &
\dtaSesMissedRemarkLog &
\dtaSesSavedHunks &
\dtaSesSavedHunkTm &
\dtaSesSavedJavaLines \\

MO\_I &
\dtaMoiComplexity &
\dtaMoiFeatureCount &
\dtaMoiMissedRemarks &
\dtaMoiMissedRemarkLog &
\dtaMoiSavedHunks &
\dtaMoiSavedHunkTm &
\dtaMoiSavedJavaLines \\

MO\_A &
\dtaMoaComplexity &
\dtaMoaFeatureCount &
\dtaMoaMissedRemarks &
\dtaMoaMissedRemarkLog &
\dtaMoaSavedHunks &
\dtaMoaSavedHunkTm &
\dtaMoaSavedJavaLines \\

RIPPER\_S\tnote{2} &
\dtaRipSComplexity &
\dtaRipSFeatureCount &
\dtaRipSMissedRemarks &
\dtaRipSMissedRemarkLog &
\dtaRipSSavedHunks &
\dtaRipSSavedHunkTm &
\dtaRipSSavedJavaLines \\

RIPPER &
\dtaRipComplexity &
\dtaRipFeatureCount &
\dtaRipMissedRemarks &
\dtaRipMissedRemarkLog &
\dtaRipSavedHunks &
\dtaRipSavedHunkTm &
\dtaRipSavedJavaLines \\
\midrule

Max.~Value\tnote{3} &
$\infty$ &
52 &
\dtaMaxMissedRemarks &
\dtaMaxMissedRemarkLog &
\dtaMaxSavedHunks &
\dtaMaxSavedHunkTm &
\dtaMaxSavedJavaLines \\ \bottomrule
\end{tabularx}
\begin{tablenotes}
\item[1] Tr.M. := trimmed mean
\item[2] During the team session it was decided to remove one further ticket from the training data. RIPPER is the rule set learned with the final data, RIPPER\_S is based on the older data and was used in the survey.
\item[3] The last row shows the total count / maximum possible value for the respective objective.
\end{tablenotes}
\end{threeparttable}
\end{table}

\begin{figure}
\begin{lstlisting}[basicstyle=\small,language=ruleset,breaklines=true]
skip when one of
  (changetype == 'DELETE')
  or (isNodeModules == 'true')
  or (packageAndImportOnly == 'true')
  or (whitespaceOnly == 'true')
  or (changeInHunkSize >= -0.5 and hunkCountInFile >= 147.5)
  or (filetype == 'jav')
  or (binary == 'true')
  or (fileCountInCommit >= 55.5 and hunkCountInCommit >= 2560.0)
  or (fileCountInCommit >= 274.0)
  or (fileCountInCommit >= 11.5 and srcdir == 'testdata')
  or (gitSimilarity >= 98.5)
  or (project == 'UnitTestRunner')
  or (project == 'TestPlugins')
  or (commitsSinceLastRemarkInFile >= 78.0)
  or (newLineCountInFile >= 12170.0)
  or (entropyCbMed <= 0.0919)
  or (visibilityChangeOnly == 'true')
\end{lstlisting}
\caption{Ruleset SESSION, \ie the ruleset that was used in the discussion with the development team. It is based on MO\_I.}
\label{fig:ruleSetSession}
\end{figure}

Figure~\ref{fig:resultsTrainingSet} also allows an estimate of the hardness of the mining task for the objectives. Without missing remarks, the best found model can save the review of \dtaNoMissSavedHunksPercent{}\% of the records. But practically more relevant are the relative numbers of saved Java lines and the trimmed mean of saved records per ticket. Both are much lower (\dtaNoMissSavedJavaLinesPercent{}\% and \dtaNoMissSavedHunksTrimmedMeanPercent{}\%), indicating that the good numbers are due to rare events: tickets with a large number of changes to non-source files. Also, all the stated numbers are optimistic, as they are evaluated on the training set. More realistic results from unseen data will follow in Section~\ref{sec:performanceUnseenData}.

All selected rulesets can be found in the online material~\citep{baum2018predictionOnlineMaterials}. To give an impression of the found rulesets, Figure~\ref{fig:ruleSetSession} shows the ruleset SESSION. It is the least complex of the five rulesets, and similar to MO\_I. Abstracting its specific contents a bit, it contains the following groups:
\begin{itemize}
\item Derived or too low-level changes, \eg changes in imports or generated code\footnote{The team uses the file extension ``jav'' to denote generated Java code.},
\item Likely systematic changes, \eg very large commits\footnote{Even when a large commit does not contain systematic changes only, the chance of finding problems in it might be lower.} or additions in files with many changes,
\item Changes that are low-risk due to previous checks (compiler, CI server, \dots), \eg deletions or pure whitespace changes,
\item Files that are empirically low-risk because there was no remark for a long time,
\item Changes that are low-risk because they concern non-production code, \eg the ``UnitTestRunner'' project, and
\item Very non-surprising changes, \ie changes with a low entropy compared to the rest of the codebase.
\end{itemize} 

The three rules that are responsible for most of the savings are \texttt{packageAnd\-ImportOnly == 'true'}, \texttt{whitespaceOnly == 'true'}, and \texttt{binary == 'true'}.  When using only these three rules, the trimmed mean of saved hunks per ticket is 8.5 records/ticket. This is a share of 84\% of the value for the whole SESSION ruleset. The most influential single rule is \texttt{packageAndImportOnly == 'true'}, with savings of 5.4 records/ticket.

\rqAnswer{RQ$_{3.2}$}{On the training set, the results of the multi-objective rule mining algorithm are better than those obtained with RIPPER. The results obtained with user interaction are better in the regions of the objective space that were focused on by the user, \ie complexity and broad ticket coverage. Figure~\ref{fig:ruleSetSession} shows the ruleset that came out of this interaction between mining tool and domain experts. 84\% of the savings of this ruleset are due to three simple syntactic rules.}

\subsection{Developers' Opinion on the Rules}
\label{sec:developersOpinion}

The requirements survey (Section~\ref{sec:characteristicsOfTargetModel}) revealed that the developers prefer to check the rules before using them. Therefore, the development team and the first author discussed the ruleset SESSION in a joint session. Before that session, we performed a survey to gather a subjective ranking for the four other rulesets. Each rule was printed on a separate sheet of paper, and the participants were asked to rate it on a scale from -5 (extremely bad) to 5 (extremely good). They were also asked to give reasons for their rating. The sheets were shuffled before handing them out and did not mention how the rules were obtained. A total of 14 developers filled out this survey and took part in the discussion.

\begin{table}
	\centering
	\caption{Survey Results for the Subjective Quality of the Mined Rulesets. All ratings are on a scale from -5 (extremely bad) over 0 (neutral) to 5 (extremely good). Rows are ordered by mean rating.}
	\label{tab:ruleSetQualityRatings}
		\begin{tabular}{lrrrr}
				  \toprule
		& \multicolumn{4}{c}{Rating} \\ \cmidrule(r){2-5}
		Ruleset & Mean & Median & Min. & Max. \\ \midrule
MO\_I     & -0.21 & 0 & -5 & 3 \\
MO\_A     & -1.08 & -1 & -5 & 3 \\
RIPPER\_S & -3.25 & -5 & -5 & 0 \\ \bottomrule
    \end{tabular}
\end{table}

Table~\ref{tab:ruleSetQualityRatings} shows that the team members considered the ruleset MO\_I to be the best, and RIPPER as worst. But even the best ruleset has a negative mean rating (-0.21). To shed light on the reasons, we analyze the textual comments from the survey and the audio recording of the discussion.

Especially the RIPPER results were criticized for being hard to understand and containing partial rules that looked nonsensical.
Large size, use of negation and rules with many numerical thresholds were especially detrimental to understanding. Often, this led to worse ratings, but sometimes the rulesets were still rated neutrally based on trust that ``there will be something good in there''.
Contrary to the other rulesets, two participants criticized MO\_I for not filtering out further change parts.

For MO\_A and MO\_I, the criticism was more towards specific rules and features. Often rules were regarded as not explicit enough, \ie they left a theoretical chance of defects slipping through. Further conditions were suggested to reduce this chance. We will discuss this problem further in Section~\ref{sec:qualitativeFeedbackFinal}. Very coarse rules that lead to the non-review of whole files or commits were also criticized. To the developers' intuition, these rules and the features used in them missed a strong link to the importance of the changes for review. Most criticized for missing this link were the time/day features. The latter was also disapproved for providing the opportunity to ``game the system'', \eg when waiting for a specific day to avoid a review. Single feature rules for example based on whitespace or entropy were instead praised as intuitive.

We could observe two opposing points of view among the company's developers. Some agreed with the point of view taken in the current article: Leaving out parts during review is a cost-benefit-tradeoff, and the costs and benefits can be derived from empirical data. Others took on a point of view that opposed leaving out change parts in reviews. The latter group often argued with the theoretical possibility of defects, no matter how small the empirical risk. Instead of shrinking the review scope, the reviewers should take more time for reviews and make pauses to avoid overload. Looking again at the results of the survey on the requirement importance (Section~\ref{sec:characteristicsOfTargetModel}), the vast spread of responses for leaving out many change parts could be explained by these opposing views.

Based on the controversial discussion in the development team, the team decided to implement the skipping rules in its review process in a limited way for now: The classification and reasons for the classification of every change part shall be visualized, but they should not be left out entirely in an automated way.

\rqAnswer{RQ$_{3.3}$}{The ruleset obtained by our interactive multi-objective approach (MO\_I) was rated best by the developers. The ruleset from RIPPER was rated worst. But even MO\_I was rated only neutral on average. Likely reasons are a rejection of specific contained rules and a general opposition to the idea of leaving out change parts in reviews by some developers.}

\subsection{Performance on Unseen Data}
\label{sec:performanceUnseenData}

\begin{figure}
	\centering
		\includegraphics[height=0.92\textheight]{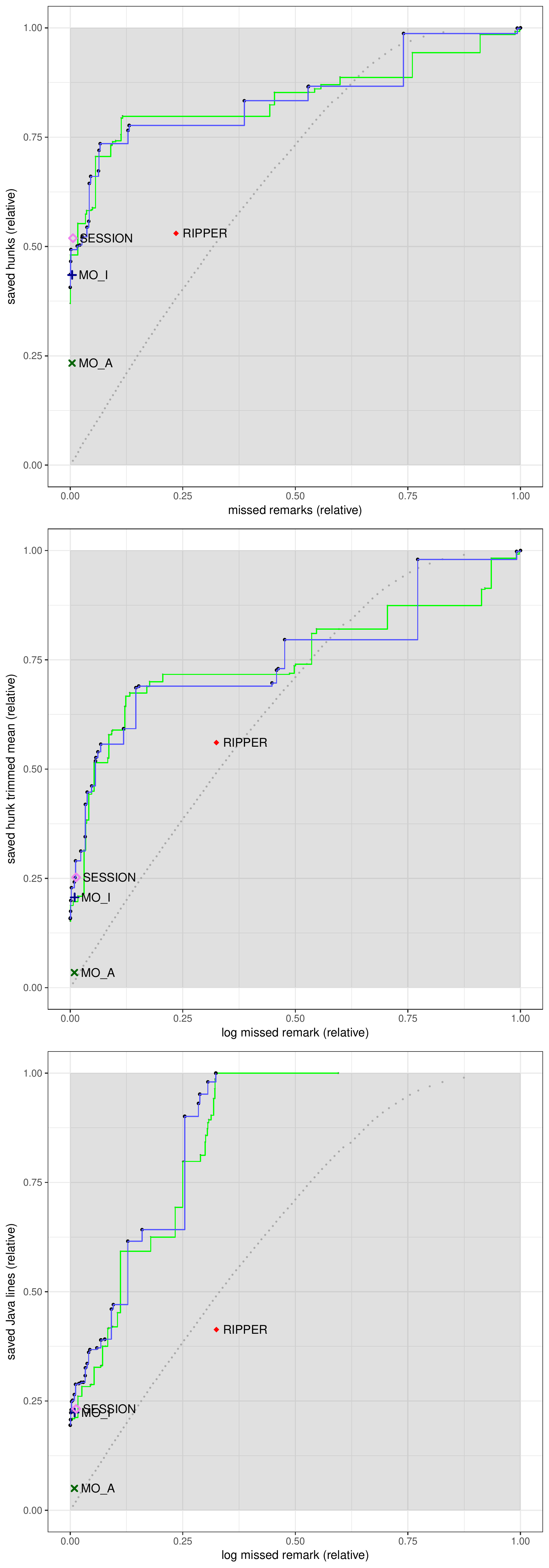}
	\caption{Pareto fronts and selected rulesets, evaluated on the unseen test data. The plots show two-dimensional projections from the seven-dimensional objective space. The gray dots show the baseline performance of leaving out a certain percentage of records per ticket; each dot corresponds to a percentage value, with results averaged over 100 random seeds.}
	\label{fig:resultsTestSet}
\end{figure}

After creating and discussing the results based on the training data, we waited until mid-November 2018 and extracted the new data that had accumulated in these 3.5 months. This test set contains data from \dtaTicketCountTestdata{} tickets. We applied the same filtering as for the training data and re-evaluated the found rulesets on the test data. Figure~\ref{fig:resultsTestSet} depicts the results in objective space. Table~\ref{tab:objectiveValuesForSelectionTestSet} shows the objective values for the five picked rulesets, and Table~\ref{tab:relativeResultsForSelectionTestSet} reframes the results for savings and missed remarks as relative values.

As expected, the results on the unseen test data are mostly worse than on the training data. Still, the only ruleset that shows a strong degradation in performance is MO\_A: The saved hunk count and the saved Java source line count plummeted. The trimmed mean of saved hunks was already low for this ruleset on the training data, indicating that the change in performance might be due to focusing too much on large but rare events. A likely cause is that MO\_A is the only ruleset whose creation did not employ an automated or interactive mechanism against this problem.
When comparing the selected rules to each other, not much has changed: RIPPER is still dominated, and the number of missed remarks is still an order of magnitude larger for it.

\begin{table}[tb!]
	\centering
	\caption{Objective Values for the Selected Rulesets on the Unseen Test Data (see Table~\ref{tab:objectiveValuesForSelection} for further descriptions).}
	\label{tab:objectiveValuesForSelectionTestSet}
		\begin{tabularx}{\textwidth}{Xrrrrrrr}
\toprule
 &
\multicolumn{4}{c}{Objectives to Minimize} &
\multicolumn{3}{c}{Objectives to Maximize} \\ \cmidrule(r){2-5} \cmidrule(r){6-8}

Ruleset &
\parbox{1.0cm}{\raggedleft Compl\-exity} &
\parbox{1.0cm}{\raggedleft Feature\\Count} &
\parbox{1.1cm}{\raggedleft Missed\\Remark\\Count} &
\parbox{1.0cm}{\raggedleft Log-Transf.\\Missed\\Remarks} &
\parbox{1.0cm}{\raggedleft Saved\\Record\\Count} &
\parbox{1.0cm}{\raggedleft Tr.M.\\Saved\\Records Per\\Ticket\\} &
\parbox{1.0cm}{\raggedleft Saved\\LOC\\in Java\\Files} \\ \midrule

SESSION &
\dtaSesUComplexity &
\dtaSesUFeatureCount &
\dtaSesUMissedRemarks &
\dtaSesUMissedRemarkLog &
\dtaSesUSavedHunks &
\dtaSesUSavedHunkTm &
\dtaSesUSavedJavaLines \\

MO\_I &
\dtaMoiUComplexity &
\dtaMoiUFeatureCount &
\dtaMoiUMissedRemarks &
\dtaMoiUMissedRemarkLog &
\dtaMoiUSavedHunks &
\dtaMoiUSavedHunkTm &
\dtaMoiUSavedJavaLines \\

MO\_A &
\dtaMoaUComplexity &
\dtaMoaUFeatureCount &
\dtaMoaUMissedRemarks &
\dtaMoaUMissedRemarkLog &
\dtaMoaUSavedHunks &
\dtaMoaUSavedHunkTm &
\dtaMoaUSavedJavaLines \\

RIPPER\_S &
\dtaRipSUComplexity &
\dtaRipSUFeatureCount &
\dtaRipSUMissedRemarks &
\dtaRipSUMissedRemarkLog &
\dtaRipSUSavedHunks &
\dtaRipSUSavedHunkTm &
\dtaRipSUSavedJavaLines \\

RIPPER &
\dtaRipUComplexity &
\dtaRipUFeatureCount &
\dtaRipUMissedRemarks &
\dtaRipUMissedRemarkLog &
\dtaRipUSavedHunks &
\dtaRipUSavedHunkTm &
\dtaRipUSavedJavaLines \\
\midrule

Max.~Value &
$\infty$ &
52 &
\dtaMaxUMissedRemarks &
\dtaMaxUMissedRemarkLog &
\dtaMaxUSavedHunks &
\dtaMaxUSavedHunkTm &
\dtaMaxUSavedJavaLines \\ \bottomrule
\end{tabularx}
\end{table}

\begin{table}[tb!]
	\centering
	\caption{Relative Objective Values (\ie Percentages of the Maximum) for the Selected Rulesets on the Unseen Test Data. Tr.M. := trimmed mean}
	\label{tab:relativeResultsForSelectionTestSet}
		\begin{tabularx}{\textwidth}{Xrrrrr}
\toprule
 &
\multicolumn{2}{c}{Objectives to Minimize} &
\multicolumn{3}{c}{Objectives to Maximize} \\ \cmidrule(r){2-3} \cmidrule(r){4-6}

Ruleset &
\parbox{1.1cm}{\raggedleft Missed\\Remark\\Count} &
\parbox{1.0cm}{\raggedleft Log-Transf.\\Missed\\Remarks} &
\parbox{1.0cm}{\raggedleft Saved\\Record\\Count} &
\parbox{1.0cm}{\raggedleft Tr.M.\\Saved\\Records Per\\Ticket\\} &
\parbox{1.0cm}{\raggedleft Saved\\LOC\\in Java\\Files} \\ \midrule

SESSION &
\dtaSesURelMissedRemarks\% &
\dtaSesURelMissedRemarkLog\% &
\dtaSesURelSavedHunks\% &
\dtaSesURelSavedHunkTm\% &
\dtaSesURelSavedJavaLines\% \\

MO\_I &
\dtaMoiURelMissedRemarks\% &
\dtaMoiURelMissedRemarkLog\% &
\dtaMoiURelSavedHunks\% &
\dtaMoiURelSavedHunkTm\% &
\dtaMoiURelSavedJavaLines\% \\

MO\_A &
\dtaMoaURelMissedRemarks\% &
\dtaMoaURelMissedRemarkLog\% &
\dtaMoaURelSavedHunks\% &
\dtaMoaURelSavedHunkTm\% &
\dtaMoaURelSavedJavaLines\% \\

RIPPER\_S &
\dtaRipSURelMissedRemarks\% &
\dtaRipSURelMissedRemarkLog\% &
\dtaRipSURelSavedHunks\% &
\dtaRipSURelSavedHunkTm\% &
\dtaRipSURelSavedJavaLines\% \\

RIPPER &
\dtaRipURelMissedRemarks\% &
\dtaRipURelMissedRemarkLog\% &
\dtaRipURelSavedHunks\% &
\dtaRipURelSavedHunkTm\% &
\dtaRipURelSavedJavaLines\% \\
\bottomrule
\end{tabularx}
\end{table}

For the SESSION ruleset, our mechanism counts 12 remarks as missed. We analyzed these in detail:
In six cases the remark would have been found anyway, \ie they are false positives.
For two of these false positives, the remark is real but not all triggers were identified, and the other four were follow-up changes and no real remarks.
This leaves six cases in which a real remark would have been missed.
Of these, two are negligible improvements. Another one is a build script maintainability issue that would have been missed because the respective commit was larger than 274 files. The most severe misses are three customer-facing documentation issues/typos that would have been missed because the file containing the documentation became longer than 12170 lines. These misses echo the developers' criticism of coarse rules that miss a strong link to the importance of the changes for review (Section~\ref{sec:developersOpinion}).

We now look at the objective values for the ruleset SESSION. It allows skipping of \dtaSesURelSavedHunkTm\% of the records per ticket (trimmed mean) and of \dtaSesURelSavedJavaLines\% of Java source code lines in reviews. According to our mechanism, this skipping will lead to non-detection of \dtaSesURelMissedRemarks\% of the review remarks (resp. \dtaSesURelMissedRemarkLog\% when using the log-transformed value). Based on the raw objective values, approximations for profit can be calculated. As motivated in Section~\ref{sec:targetMetric}, these values depend on a cost factor that determines how much more costly missing a remark in a review is compared to reviewing a change part. A break-even value that determines when the profit will become positive can be calculated. When using the log-transformed remark count and the trimmed mean of saved records per ticket, this break-even value is \dtaSesUBreakEvenRecord{}, \ie using the ruleset has a positive profit as long as missing a review remark is less than \dtaSesUBreakEvenRecord{} times as costly as reviewing a record. When using Java lines of code as the approximation of effort instead, the profit is positive as long as missing a review remark is less than \dtaSesUBreakEvenLoc{} times as costly as reviewing a line of Java code.

The three simple syntactic rules that were responsible for most of the savings for the SESSION ruleset on the training data are still responsible for most of the savings on the test data. \texttt{packageAnd\-ImportOnly == 'true'} is still most influential, and \texttt{whitespaceOnly == 'true'} and \texttt{binary == 'true'} have switched places, with binary being second. When using only these three rules, the trimmed mean of saved hunks per ticket is 7.8 records/ticket. This is a share of 82\% of the value for the whole SESSION ruleset.

\rqAnswer{RQ$_{3.4}$}{The ruleset SESSION, and the closely related MO\_I, both perform well on the unseen test data. They allow savings of more than one-fifth of reviewed records and Java LOC. In turn, they lead to about one percent of missed remarks. The RIPPER ruleset leads to much higher numbers of missed remarks. MO\_A filters out very little on the test data.}

\section{Discussion and Future Work}
\label{sec:futureWork}

Next, we discuss open problems of our approach, before we propose ideas for future work in the section afterward.

\subsection{Discussion}
\label{sec:qualitativeFeedbackFinal}

We could successfully extract data and mine useful rules, but some problems were not yet solved to our satisfaction:

\noindent\textbf{Noise.}
Section~\ref{sec:iterativeImprovement} describes a significant amount of noise found in the data, due to failed assumptions of our tracing algorithm. Other problems are caused by non-obedience to the development process, \eg when remarks are not fixed in the right ticket or the ticket state is not changed correctly. Noise is a significant problem: The most relevant target for the mining process is the profit of using the ruleset. The cost of missing a remark is magnitudes larger than that of reviewing a change part. Therefore, the profit metric behaves similarly to precision, which is known to be vulnerable to noise~\citep{menzies2007problems}. Many potentially interesting rules might be obscured by the noise in the data.

\noindent\textbf{Unknown individual costs.}
The cost per individual missed remark is not known. This can lead to wrong incentives for the mining algorithm, which might opt for a rule that leads to the oversight of one critical error instead of another that misses two trivial issues. The same problem exists for the cost of reviewing change parts, but this cost can at least be approximated for example by the number of lines of code.

\noindent\textbf{Reliance on Occam's razor.}
Ruleset complexity is one of the algorithm's objectives for two reasons: (1)~Simpler rules are easier to understand. (2)~There is the hope that of two competing and otherwise similar rules, the simpler one is the better model of reality (Occam's razor). Albeit useful, Occam's razor is only a heuristic that fails sometimes. An example that was criticized by some of the company's developers illustrates this: The algorithm selected a rule to ignore whitespace-only changes. This rule is straightforward, and it is also highly successful, as there are many whitespace-only changes in Java files that do not lead to review remarks. But in other file types, whitespace changes are much more dangerous and remark-prone. When changes in these files occur rarely, the ``ignore whitespace-only changes'' rule will seem better than the ``ignore whitespace-only changes in Java files'' rule, albeit the latter better represents the underlying domain.

\noindent\textbf{Reliance on historical data.}
A general problem of every empirical approach for rule creation is that the rules are built for past data. Changes in the project structure, development processes, or other areas can make found rules obsolete. One example is the growth in file size that led to the missed remarks described in Section~\ref{sec:performanceUnseenData}. Another example is the ``isNodeModules'' rule in MO\_I, MO\_A, and SESSION, which relates to the practice of committing certain libraries to the SCM. This practice was stopped by the development team so that the rule will not lead to savings in the future. Besides input from domain experts, regular re-training can help to reduce this problem.

\noindent\textbf{Amplification of reviewers' weaknesses and self-fulfilling prophecies.}
If a change part does not trigger review remarks, this can have two causes: There is no problem, or the reviewer did not find the problem. The latter case is disputed: When there are changes in which the reviewers consistently overlook problems, this weakness of the reviewers might be institutionalized as a skipping rule. This rule deprives the reviewers of the remaining small chance to find the issue. The problem could be reduced by combining a model for skipping with the SRK model discussed in Section~\ref{sec:classificationImproveReview}, or with a general defect prediction model. A variant of the problem occurs when data from a team in which skipping rules are in use is used to re-evaluate the rules: As the reviewers did not see the change parts that are matched by the rules, they probably left few remarks for them. So, the existing rules will look good no matter how good they really are -- a self-fulfilling prophecy.

\noindent\textbf{Much ado about .. not much.}
A large share of the savings of the SESSION and MO\_I rulesets are due to simple syntactic rules. This leaves the impression that similar benefits for the company could have been reaped in a more lightweight way, without going through the hassle of extracting the remark data and performing a large-scale data mining study. We can only guess whether less noisy data would allow the extraction of less obvious high-impact rules or whether there is just nothing to be found. Please note that the criticism of the high effort applies mostly to the action research view of our study; we still consider it worthwhile to gather and analyze empirical data in detail from a basic research perspective.

Most of the mentioned problems are not unique to our study; they are shared by many other repository mining studies. This might contribute to their still relatively low use in industry.

\subsection{Implications and Future Work}

We will now discuss the implications of our study and resulting ideas for future work. 
For practitioners, our study implies that they should use a mechanism to leave out unimportant change parts in reviews. Reducing the review scope by more than 20\% while missing out very few remarks is an opportunity that should not be left untapped. But when it comes to how to establish the model underlying this mechanism, we cannot whole-heartedly recommend practitioners to use repository mining as we did. Most of the savings we found are due to simple syntactic rules, and in an environment where `time is money', it seems to be advisable to reap these low-hanging fruits and wait for research to come up with improved approaches.

For research, in turn, our study establishes that review remark prediction is a distinct application area in repository mining and that there is much open work in this area. We only worked on one of the possibilities outlined in Section~\ref{sec:possibilitiesChangePartImportance}, and further delving into the distinction between knowledge-based and rule-based cognitive processing in reviews could lead to interesting results. Like for every empirical study, replication in another setting could strengthen the foundation of the results. But before that, it seems advisable to tackle the open problems of our approach, most importantly the reduction of noise in the data.

Future work should analyze whether extracting the review remarks from the review tool repository instead of the SCM (Section~\ref{sec:selectingDataSource}) reduces noise and leads to better results. The concept of scopes plays a major role in our tracing algorithm (Section~\ref{sec:tracingAlgorithm}), but so far it is mainly based on intuition. Is enlarging the scope as we do it the best choice, or should, for example, physical closeness play a larger role?
Being a trigger could also be based on navigation patterns and relatedness~\citep{baum2017icsme}, so that it could be better to regard a change in a calling method as a more likely trigger than a change in the same file but far away. As some of the problems with tracing were caused by mixing follow-up changes with real review remarks, better identification of follow-up changes could lead to better results, and exploiting existing refactoring detection approaches (\eg \cite{Prete2010,Biegel2011}) could be a first step in this regard.
 More generally, the identification of the most likely triggers from the set of potential triggers could be regarded as a mining problem itself. When seen in this way, things like the similarity of a remark to the potential trigger could also be influencing factors.

As motivated in Section~\ref{sec:classificationImproveReview}, a mechanism to shrink code changes should ensure that the shrunk change is still understandable. We have not studied this question in much detail, and future work should close this gap. It could, for example, be possible to devise an ``understandability hull'' operator that takes a shrunk changeset and adds change parts deemed necessary for understanding. The review tool would then first determine the change parts needed as review triggers with the mechanisms described in this article and then apply the understandability hull to ensure good understandability of the shrunk changeset.


Instead of trying to determine the importance of change parts automatically, one could build on the author to annotate its changes~\citep{DAmbros2010}. We want to lessen the burden on the author and not mandatorily demand such additional work, but giving the author the option to add information about importance could augment our approach.

Other parts of the development process and the tools used in them influence the potential to leave out change parts. The set of defects that can be found in reviews depends on the checks that were performed before the review. For example, knowing that the code compiled successfully indicates that the risk of problems caused by changes in the package declaration is low. Similarly, future advanced static analysis tools could lead to further chances to leave out change parts in reviews.

There are cases in which a higher abstraction level than low-level SCM changes is most relevant for review. For deletions, for example, it is mostly relevant what file or functionality was deleted, and not the exact code that was contained in the deleted portions. Having automatically generated summaries of the change can help to spot problems at the optimal abstraction level and allows skipping more low-level changes.

In the discussion, the company developers noted that the way of visualizing the classification of the change parts is important. The change parts should still be openable on demand, and users want to configure what they see.
Visualization of classification results could be improved with the ideas of~\citet{Murphy-Hill2010} for interactive ambient visualization.

Besides improving the approach, it also needs more empirical validation. This encompasses the intended effects, \eg how much effort is saved by leaving out the change parts, but also unintended side-effects, \eg whether leaving out change parts hampers the spreading of knowledge in teams.


Our study also has implications for other repository mining studies. Based on our results, we encourage other authors to try multi-objective and interactive approaches to data mining when domain feedback and acceptance are essential. The use of trimmed means over some sensible aggregate (like tickets in our case) could also be beneficial in other cases. And given that many of the more successful features in our model were found with an inductive approach, the use of systematic theory-generating methodologies for feature creation seems advisable for other studies, too.

\section{Threats to Validity}
\label{sec:threats}

Several limitations apply to our study. One of the central limitations stems from being a single case study, which makes the generalizability of our results to other contexts hard to judge. This applies to the mined data and found rules as well as to the results of the surveys and discussion sessions. As an example, the high savings due to leaving out changes in import statements are probably influenced by the use of version numbers in some package names and by a positive attitude towards refactorings, both of which might not exist in other contexts. To allow others to assess further similarities and differences to their situation, we describe the context in Section~\ref{sec:caseStudySetting}.

Another limitation is that we did not measure the final outcomes of introducing the model in practice. It is well-established that a smaller review scope leads to better results~\citep{baum2018orderingExperiment} and our results indicate that applying the model will safe effort, but we do not measure the construct ``effort'' explicitly. The validity of many other of the measured constructs depends on the correctness of the mining and extraction programs. We tested many parts of the source code with automated unit tests. For central parts, especially in the tracing and mining algorithms, we tested to 100\% branch coverage and used mutation testing with PIT\footnote{http://pitest.org} to ensure adequate coverage. But as a heuristic approach, the mechanism to count the number of missed review remarks can still fail. It might do so in two ways: identifying too many or too few missed remarks. By manually checking a sample of the remarks, we were able to assess the ``too many'' case. But the ``too few'' case, \eg review remarks that were fixed in a commit that was not correctly labeled with the ticket number, was only assessed qualitatively.

A usual practice in data mining studies is to use repeated cross-validation to statistically judge the quality of the results. This is not possible with our interactive approach. Instead, we use a training set and a separate test set. Having only one test set leads to less reliable results. On the other hand, using a test set collected after mining is a very realistic way of performance evaluation.
To put our interactive, multi-objective mining approach in context, we compare it to RIPPER as a baseline. In defect prediction studies, other mining approaches outperform RIPPER~\citep{tantithamthavorn2018impactOpti}. But these approaches, like random forests, lead to opaque models that do not meet the team's requirements. Therefore, a rule mining approach is an adequate baseline.

Various threats apply to the opinions gathered from the developers. In the interactive evaluations sessions, the participants needed more time than initially expected to understand the system, as they were overwhelmed by the current UI of the system. This limited the amount of feedback that could be gathered in the available time.
The opinions on the mined rulesets were gathered directly before the discussion session. As some of the presented rules were clearly sub-optimal, this might have put some developers in a negative mood for the discussion. Another threat here is reactivity, in particular, because the researcher presenting the approach and the rules was also a colleague.

For the analysis of the qualitative data, we used best practices from qualitative data analysis, like transcribing the audio recordings and performing several passes of open coding. Still, having only one coder amplifies the risk of researcher bias.

To allow others to judge this and other potential biases, we make the session transcripts available as part of the online appendix. All other raw data is also made available, albeit we anonymized the dataset to protect confidential company data. With an interactive data mining approach like ours, it is not sufficient to make the data mining algorithms and scripts available to allow others to double-check or recreate the results. Therefore, we also provide logs of the interactive sessions as an audit trail.

\section{Related Work}
\label{sec:relatedWork}

The notion of cognitive support code review tools was proposed by~\citet{baum2016profes}. It is based on the hypothesis that reducing the cognitive load~\citep{paas1994instructional} of reviewers will improve review performance. An experimental assessment of aspects of this hypothesis can be found in \citet{baum2018orderingExperiment}. Positive effects of reducing the review size are also discussed in other works (\eg~\citet{Raz1997,rigby2012contemporary,Rigby2013}). In Section~\ref{sec:classificationImproveReview} we also use another theory from cognitive psychology, the SRK taxonomy that divides human cognitive processing into three modes that differ in the degree of automation and efficiency of processing~\citep{rasmussen1983skills}. A general analysis of the utility of these and other theories for software engineering research is performed by~\citet{Walenstein2003}.

SZZ~\citep{Sliwerski2005} is the standard algorithm for tracing defect fixes to the introducing changes in defect prediction studies. \citet{kim2006automatic} proposed a variant of this algorithm that reduces the noise in the extracted data. Like our tracing algorithm for review remarks, they skip changes that cannot be a cause/trigger. The problem of noise in defect prediction is studied by \citet{kim2011dealing}.

Review remark prediction is not the same as defect prediction: A change part might contain a defect that does not lead to a review remark, and the different assumptions during tracing also lead to differences in the raw data (see Section~\ref{sec:comparisonToSzz}). Still, review remark prediction and defect prediction are related. Defect prediction and repository mining are vast research areas, so we can not provide an exhaustive overview of the literature.

Our algorithm classifies at the level of change parts. Most defect prediction studies work at a coarser granularity, \eg predicting defect proneness for whole changes (also called ``just-in-time'' prediction) or for methods, files or components.
An early study to predict the risk of a software change was performed by \citet{mockus2000predicting} at Bell Labs. \citet{Shihab2012} use input from industrial developers to assess the risk of changes. One of the problems of our study is that we consider the severity of the review remarks only to a limited degree. \citet{shihab2011high} tackle a similar problem for just-in-time defect prediction and study ``high impact'' fixes. Results of just-in-time prediction at Google were disappointing \citep{lewis2013does}, but \citet{tan2015online}'s case study at Cisco and especially \citet{nayrolles2018clever}'s case study at Ubisoft show that an elaborated approach can bring just-in-time defect prediction to industrial usefulness. Further examples of defect prediction studies with change granularity are \citet{Kim2008}, \citet{kamei2013large}, \citet{shivaji2013reducing}, and \citet{soltanifar2016predicting}.
\citet{Ray2015} show that the entropy of code is related to its defect density and go down to the level of single lines.
Results of \citet{pascarella2018re}, when re-evaluating a study by \citet{Giger2012}, indicate that the current state of the art of method-level defect prediction does not lead to results that are satisfying in practice. \citet{shippey2015exploiting} analyzes the relation of AST patterns to defective and non-defective methods. He finds associations for defective methods, but not for non-defective methods.

We argue that the meta-parameters of many data mining techniques are hard to interpret for domain experts. Other researchers (\eg \citet{tantithamthavorn2018impactOpti}) use automatic approaches to optimize these meta-parameters.
\citet{arisholm2010systematic} criticize that often used evaluation measures like precision and recall do not directly relate to cost-effectiveness. As \citet{kamei2010revisiting} show, effort-aware evaluation of models leads to different conclusions. A benefit of our approach is that it is easy to include domain-specific evaluation measures.
Our decision to go for a multi-objective mining technique was encouraged by \citet{fu2018building}'s promising results with DART, which can be regarded as an ensemble of rules created with a multi-objective approach.
Further studies on defect prediction are surveyed by \citet{hall2012systematic}, \citet{Radjenovic2013}, and \citet{malhotra2015systematic}.

Review remarks can relate not only to defects but also to other quality problems. These, too, can be found with data mining, for example like \citet{Fontana2015} who predict code smells.

In our study, we focus on ways to support the reviewer. Another avenue to improve code reviews is to create automated review agents~\citep{Chan2001}, \ie programs that directly create review remarks. Just-in-time defect prediction is one possibility to create such automated reviewers \citep{rosen2015commit,fejzer2015supporting,nayrolles2018clever}. Another possibility is to use results from static analysis \citep{Brothers1990,Belli1997,Farchi2006,henley2018cfar}.
Automated reviewers can already work during check-in \citep{Tarvo2013,Baum2015} or even in parallel to development \citep{madhavan2007predicting}.
In Section~\ref{sec:classificationImproveReview} we discuss sorting change parts by their importance for review. Similar ideas have been studied by other researchers, \eg by \citet{lumpe2012learning}.
The idea to focus reviewing on ``sections where finding defects is really worthwhile'' can already be found in \citet[p.~74]{Gilb1993}, albeit they relate more to the severity of defects and not to the probability of finding them. \citet{begel2018eye} study eye movements of reviewers and find that developers focus on relevant portions instead of the entire text, which supports our idea of automatically classifying parts as irrelevant. \citet{huang2018salient} propose a machine learning approach to identify the `salient class' in a change, \ie the class whose change is at the core of the commit. This could be a further criterion to rate the importance of code changes for review.
The idea to leave out lines of lesser importance to improve understanding has been studied in the context of end-user programming by \citet{Athreya2014}.

Our approach is bottom-up and empirically determines change parts that are irrelevant for review. Other researchers instead start by assuming that certain types of changes are irrelevant. Refactorings are one such type, and \citet{ge2017} propose to use refactoring detection in code reviews. \citet{Thangthumachit2011} use refactoring detection to improve the understandability of source code changes. Going beyond refactorings, \citet{Kim2013} describe how to discover systematic code changes and \citet{Zhang2015} propose a tool to interactively match systematic changes in code reviews.
Simple syntactic rules (\eg whitespace and import statements) are responsible for a large part of the savings with the selected rulesets. Similar results could be obtained by using semantic or otherwise improved diff algorithms \citep{Apiwattanapong2004,fluri2007change,Yu2011,Kawrykow2011}.

We study the importance of change parts for review. Other aspects of review data have also been assessed in data mining studies: For example, \citet{gerede2018will} predict at the coarse granularity of whole patches whether they will lead to review remarks. \citet{Padberg2004} predict the defect content after reviews and \citet{kononenko2015investigating} study factors that influence review quality in Firefox subprojects. \citet{Bird} have deployed a code review analytics platform at Microsoft, which could also be useful to determine irrelevant change parts. It was used by \citet{Bosu} to develop a model to distinguish useful review remarks from less useful ones (\ie noise).

We use a multi-objective rule mining meta-heuristic and domain feedback to create comprehensible rules. With the same goal, \citet{vandecruys2008mining} use the AntMiner+ meta-heuristic for mining. Other researchers use a two-step process: First, a black box model is learned, and as a second step this model is transformed into a comprehensible model \citep{moeyersoms2015comprehensible}. Still another approach is to create explanations for the model's decision upon request \citep{tan2015online,dam2018explainable}.
Multi-objective meta-heuristics were also used in other software engineering studies, often based on genetic algorithms. They are used for the model creation itself \citep{de2010symbolic,chen2018multi}, but also for feature selection \citep{chen2017applying} or model refinement \citep{canfora2015defect,xia2016collective,ryu2016effective}. There are also studies that explicitly incorporate iterative human feedback, \eg to determine patterns for implicit coding rules \citep{matsumura2002detection}, common defects \citep{Williams2005} and requirements tracing links \citep{hayes2006advancing}. 

\section{Conclusion}
\label{sec:conclusion}

We performed a case study in a medium-sized software company. In it, we analyze ways to use repository mining to identify the importance of change parts for code review and to improve code review efficiency based on this information.
We propose that information on change part importance can help the reviewer to reach code review goals more efficiently by: (1)~Leaving out change parts that neither trigger review remarks nor are needed for understanding, (2)~highlighting change parts that the reviewer should review in a higher cognitive mode than he or she normally would, or (3)~order change parts from the most likely to the least likely remark trigger, as long as this order is still easily understandable. We focus on (1) in the article and take the point of view that a slight decrease in found remarks is acceptable when offset by a large saving in effort, \ie when the overall profit is positive.
We gather data to predict change part importance from software repositories.
Review remarks can be extracted either from code review repositories or from review commits in the SCM. We focus on the latter and propose an algorithm to associate review remarks with potential triggers. The algorithm finds these triggers by tracing back in history, expanding the search scope if no matching trigger could be found with the smaller scope.
To compare that algorithm to a simpler alternative, we compare it with the SZZ tracing approach.
For \dtaSzzDiffPercent{}\% of the review remarks, using SZZ to trace remarks to triggers would lead to entirely different results. The majority of differences is caused by not enlarging the search scope when no trigger is found among the commits of the ticket under review.
We gather requirements for the prediction model from the company's development team.
The most important requirements from the team are that the model provokes as few missed review remarks as possible, does not lead to additional waiting time in the review and is transparent to and checked by the team.
To satisfy these requirements, we use an interactive, multi-objective rule mining approach. We compare it to RIPPER as a baseline.
When assuming that missing a review remark is much more costly than reviewing a change part, the results of the multi-objective rule mining algorithm are better than those obtained with RIPPER. The results obtained with user interaction are better in the regions of the objective space that were focused on by the user, \ie complexity and broad ticket coverage, and they show fewer signs of overfitting to the training set. The ruleset obtained by our interactive multi-objective approach was also rated best by the developers among the studied alternatives.
On the test set, the finally selected ruleset allows savings of \dtaSesURelSavedHunkTm\% of the reviewed change part records per ticket and of \dtaSesURelSavedJavaLines\% of the Java source lines. In turn, it leads to about one percent of missed remarks.
Despite these good results, there are several remaining problems.
Some of the developers show a general opposition to the idea of leaving out change parts in reviews. And about 80\% of the savings were due to three simple syntactic rules, which does not reap the full potential of a data mining approach. Reduction of noise in the input data could lead to better results. All in all, we hope that our work is a useful first step on review remark prediction that will be improved upon by future work.

\begin{acknowledgements}
The authors would like to thank the case company and its developers for the time and effort they donated. Furthermore, we would like to thank Eirini Ntoutsi for input on the data mining approach, Javad Ghofrani for providing his mining server and Melanie Busch and Wasja Brunotte for feedback on an article draft.
\end{acknowledgements}

\bibliographystyle{spbasic}      
\bibliography{C:/Users/ich/Documents/literatur/literatur}

\end{document}

%% file: writing_support.tex
\usepackage{amssymb}
\usepackage{xspace}
\usepackage{framed}
\usepackage{booktabs}

\newcommand{\ie}{\emph{i.e.},\xspace}
\newcommand{\eg}{\emph{e.g.},\xspace}

\definecolor{gray50}{gray}{.5}
\definecolor{gray40}{gray}{.6}
\definecolor{gray30}{gray}{.7}
\definecolor{gray20}{gray}{.8}
\definecolor{gray10}{gray}{.9}
\definecolor{gray05}{gray}{.95}

\newlength\Linewidth
\def\findlength{\setlength\Linewidth\linewidth
	\addtolength\Linewidth{-4\fboxrule}
	\addtolength\Linewidth{-3\fboxsep}
}
\newenvironment{examplebox}{\begin{center}\par\begingroup
	\setlength{\fboxsep}{5pt}\findlength
	\setbox0=\vbox\bgroup\noindent
	\hsize=0.95\linewidth
	\begin{minipage}{0.95\linewidth}\normalsize}
	{\end{minipage}\egroup
	\textcolor{gray20}{\fboxsep1.5pt\fbox
		{\fboxsep5pt\colorbox{gray05}{\normalcolor\box0}}}
	\endgroup\par\noindent
	\normalcolor\ignorespacesafterend\end{center}}

%% file: data.tex
\newcommand{\dtaMoiComplexity}{58}
\newcommand{\dtaMoiFeatureCount}{17}
\newcommand{\dtaMoiMissedRemarks}{338}
\newcommand{\dtaMoiMissedRemarkLog}{40.6}
\newcommand{\dtaMoiSavedHunks}{384,058}
\newcommand{\dtaMoiSavedHunkTm}{8.8}
\newcommand{\dtaMoiSavedJavaLines}{292,371}
\newcommand{\dtaMoaComplexity}{184}
\newcommand{\dtaMoaFeatureCount}{24}
\newcommand{\dtaMoaMissedRemarks}{0}
\newcommand{\dtaMoaMissedRemarkLog}{0.0}
\newcommand{\dtaMoaSavedHunks}{300,266}
\newcommand{\dtaMoaSavedHunkTm}{1.2}
\newcommand{\dtaMoaSavedJavaLines}{161,920}
\newcommand{\dtaRipComplexity}{342}
\newcommand{\dtaRipFeatureCount}{25}
\newcommand{\dtaRipMissedRemarks}{12,867}
\newcommand{\dtaRipMissedRemarkLog}{2,245.3}
\newcommand{\dtaRipSavedHunks}{460,408}
\newcommand{\dtaRipSavedHunkTm}{22.1}
\newcommand{\dtaRipSavedJavaLines}{464,148}
\newcommand{\dtaRipSComplexity}{200}
\newcommand{\dtaRipSFeatureCount}{17}
\newcommand{\dtaRipSMissedRemarks}{16,806}
\newcommand{\dtaRipSMissedRemarkLog}{2,493.6}
\newcommand{\dtaRipSSavedHunks}{445,275}
\newcommand{\dtaRipSSavedHunkTm}{21.6}
\newcommand{\dtaRipSSavedJavaLines}{431,981}
\newcommand{\dtaSesComplexity}{40}
\newcommand{\dtaSesFeatureCount}{17}
\newcommand{\dtaSesMissedRemarks}{1,500}
\newcommand{\dtaSesMissedRemarkLog}{142.4}
\newcommand{\dtaSesSavedHunks}{410,974}
\newcommand{\dtaSesSavedHunkTm}{10.1}
\newcommand{\dtaSesSavedJavaLines}{303,943}
\newcommand{\dtaMoiUComplexity}{58}
\newcommand{\dtaMoiUFeatureCount}{17}
\newcommand{\dtaMoiUMissedRemarks}{8}
\newcommand{\dtaMoiUMissedRemarkLog}{3.1}
\newcommand{\dtaMoiUSavedHunks}{11,733}
\newcommand{\dtaMoiUSavedHunkTm}{7.8}
\newcommand{\dtaMoiUSavedJavaLines}{11,004}
\newcommand{\dtaMoaUComplexity}{184}
\newcommand{\dtaMoaUFeatureCount}{24}
\newcommand{\dtaMoaUMissedRemarks}{8}
\newcommand{\dtaMoaUMissedRemarkLog}{2.9}
\newcommand{\dtaMoaUSavedHunks}{6,293}
\newcommand{\dtaMoaUSavedHunkTm}{1.3}
\newcommand{\dtaMoaUSavedJavaLines}{2,469}
\newcommand{\dtaRipUComplexity}{342}
\newcommand{\dtaRipUFeatureCount}{25}
\newcommand{\dtaRipUMissedRemarks}{456}
\newcommand{\dtaRipUMissedRemarkLog}{100.9}
\newcommand{\dtaRipUSavedHunks}{14,294}
\newcommand{\dtaRipUSavedHunkTm}{21.2}
\newcommand{\dtaRipUSavedJavaLines}{20,338}
\newcommand{\dtaRipSUComplexity}{200}
\newcommand{\dtaRipSUFeatureCount}{17}
\newcommand{\dtaRipSUMissedRemarks}{539}
\newcommand{\dtaRipSUMissedRemarkLog}{110.1}
\newcommand{\dtaRipSUSavedHunks}{14,166}
\newcommand{\dtaRipSUSavedHunkTm}{19.2}
\newcommand{\dtaRipSUSavedJavaLines}{18,784}
\newcommand{\dtaSesUComplexity}{40}
\newcommand{\dtaSesUFeatureCount}{17}
\newcommand{\dtaSesUMissedRemarks}{12}
\newcommand{\dtaSesUMissedRemarkLog}{4.1}
\newcommand{\dtaSesUSavedHunks}{13,998}
\newcommand{\dtaSesUSavedHunkTm}{9.5}
\newcommand{\dtaSesUSavedJavaLines}{11,425}
\newcommand{\dtaMaxMissedRemarks}{68,960}
\newcommand{\dtaMaxMissedRemarkLog}{8,055.8}
\newcommand{\dtaMaxSavedHunks}{703,706}
\newcommand{\dtaMaxSavedHunkTm}{43.9}
\newcommand{\dtaMaxSavedJavaLines}{1,155,636}
\newcommand{\dtaMaxUMissedRemarks}{1,940}
\newcommand{\dtaMaxUMissedRemarkLog}{310.7}
\newcommand{\dtaMaxUSavedHunks}{26,968}
\newcommand{\dtaMaxUSavedHunkTm}{37.8}
\newcommand{\dtaMaxUSavedJavaLines}{49,199}
\newcommand{\dtaMoiURelMissedRemarks}{0.4}
\newcommand{\dtaMoiURelMissedRemarkLog}{1.0}
\newcommand{\dtaMoiURelSavedHunks}{43.5}
\newcommand{\dtaMoiURelSavedHunkTm}{20.7}
\newcommand{\dtaMoiURelSavedJavaLines}{22.4}
\newcommand{\dtaMoaURelMissedRemarks}{0.4}
\newcommand{\dtaMoaURelMissedRemarkLog}{0.9}
\newcommand{\dtaMoaURelSavedHunks}{23.3}
\newcommand{\dtaMoaURelSavedHunkTm}{3.5}
\newcommand{\dtaMoaURelSavedJavaLines}{5.0}
\newcommand{\dtaRipURelMissedRemarks}{23.5}
\newcommand{\dtaRipURelMissedRemarkLog}{32.5}
\newcommand{\dtaRipURelSavedHunks}{53.0}
\newcommand{\dtaRipURelSavedHunkTm}{56.1}
\newcommand{\dtaRipURelSavedJavaLines}{41.3}
\newcommand{\dtaRipSURelMissedRemarks}{27.8}
\newcommand{\dtaRipSURelMissedRemarkLog}{35.4}
\newcommand{\dtaRipSURelSavedHunks}{52.5}
\newcommand{\dtaRipSURelSavedHunkTm}{50.8}
\newcommand{\dtaRipSURelSavedJavaLines}{38.2}
\newcommand{\dtaSesURelMissedRemarks}{0.6}
\newcommand{\dtaSesURelMissedRemarkLog}{1.3}
\newcommand{\dtaSesURelSavedHunks}{51.9}
\newcommand{\dtaSesURelSavedHunkTm}{25.2}
\newcommand{\dtaSesURelSavedJavaLines}{23.2}
\newcommand{\dtaNoMissSavedHunksPercent}{46}
\newcommand{\dtaNoMissSavedHunksTrimmedMeanPercent}{10}
\newcommand{\dtaNoMissSavedJavaLinesPercent}{17}
\newcommand{\dtaSesUBreakEvenRecord}{726}
\newcommand{\dtaSesUBreakEvenLoc}{2,793}

%% file: dataSzz.tex
\newcommand{\dtaSzzTotalRemarks}{326,066}
\newcommand{\dtaSzzSameResultCount}{119,031}
\newcommand{\dtaSzzSameResultPercent}{37}
\newcommand{\dtaSzzIncompleteCount}{9,947}
\newcommand{\dtaSzzIncompletePercent}{3}
\newcommand{\dtaSzzDiffCount}{197,088}
\newcommand{\dtaSzzDiffPercent}{60}
\newcommand{\dtaSzzStuckCount}{70,582}
\newcommand{\dtaSzzStuckPercent}{22}
\newcommand{\dtaSzzScopeCount}{126,506}
\newcommand{\dtaSzzScopePercent}{39}

%% file: dataPerTicket.tex
\newcommand{\dtaAllCommitsTotal}{23,853}
\newcommand{\dtaAllCommitsPerTicketMin}{1}
\newcommand{\dtaAllCommitsPerTicketMax}{88}
\newcommand{\dtaAllCommitsPerTicketLowQuart}{2}
\newcommand{\dtaAllCommitsPerTicketMedian}{3}
\newcommand{\dtaAllCommitsPerTicketUppQuart}{5}
\newcommand{\dtaImplCommitsTotal}{15,365}
\newcommand{\dtaImplCommitsPerTicketMin}{1}
\newcommand{\dtaImplCommitsPerTicketMax}{65}
\newcommand{\dtaImplCommitsPerTicketLowQuart}{1}
\newcommand{\dtaImplCommitsPerTicketMedian}{1}
\newcommand{\dtaImplCommitsPerTicketUppQuart}{3}
\newcommand{\dtaReviewCommitsTotal}{8,488}
\newcommand{\dtaReviewCommitsPerTicketMin}{0}
\newcommand{\dtaReviewCommitsPerTicketMax}{32}
\newcommand{\dtaReviewCommitsPerTicketLowQuart}{0}
\newcommand{\dtaReviewCommitsPerTicketMedian}{1}
\newcommand{\dtaReviewCommitsPerTicketUppQuart}{2}
\newcommand{\dtaRecordsTotal}{703,706}
\newcommand{\dtaRecordsPerTicketMin}{1}
\newcommand{\dtaRecordsPerTicketMax}{53,594}
\newcommand{\dtaRecordsPerTicketLowQuart}{10}
\newcommand{\dtaRecordsPerTicketMedian}{27}
\newcommand{\dtaRecordsPerTicketUppQuart}{77}
\newcommand{\dtaRemarksTotal}{68,960}
\newcommand{\dtaRemarksPerTicketMin}{0}
\newcommand{\dtaRemarksPerTicketMax}{1,437}
\newcommand{\dtaRemarksPerTicketLowQuart}{0}
\newcommand{\dtaRemarksPerTicketMedian}{2}
\newcommand{\dtaRemarksPerTicketUppQuart}{10}
\newcommand{\dtaTicketCount}{6,005}
\newcommand{\dtaTicketCountTestdata}{311}
\newcommand{\dtaJavaRecordsTotal}{370,130}
\newcommand{\dtaJavaRecordsPerTicketMin}{0}
\newcommand{\dtaJavaRecordsPerTicketMax}{5,749}
\newcommand{\dtaJavaRecordsPerTicketLowQuart}{4}
\newcommand{\dtaJavaRecordsPerTicketMedian}{18}
\newcommand{\dtaJavaRecordsPerTicketUppQuart}{57}

\newcommand{\dtaNoTriggerRecordCount}{439,626}
\newcommand{\dtaMustRecordCount}{14,385}